\documentclass[twocolumn,preprintnumbers,amsmath,amssymb,nofootinbib
,superscriptaddress]{revtex4}
\usepackage{hyperref}
\usepackage{graphicx}
\usepackage{color}
\usepackage{xcolor}
\usepackage{comment}
\usepackage{lmodern}

\newcommand{\nonumbo}{\nonumber \\ && }
\newcommand{\backo}{\!\!\!\!\!\!\!\!\!\!}

\newcommand{\smebo}{\eea }

\newcommand{\smbbo}{\small \bea  && }
\newcommand{\bl}{\biggl(}
\newcommand{\br}{\biggr)}
\newcommand{\Tr}{\makebox{Tr}}
\newcommand{\vvx}{\vec{x}}
\newcommand{\vvy}{\vec{y}}
\newcommand{\vvr}{\vec{r}}

\newcommand{\vvX}{\vec{X}}
\newcommand{\vvq}{\vec{q}}

\newcommand{\be}{\begin{equation}}  
\newcommand{\ee}{\end{equation}}  
\newcommand{\bear}{\begin{eqnarray}}   
\newcommand{\eear}{\end{eqnarray}}  
\newcommand{\ba}{\begin{array}}  
\newcommand{\ea}{\end{array}}

\usepackage{diagbox}

\definecolor{rossoCP3}{cmyk}{0,.88,.77,.40}
\definecolor{blueRef}{rgb}{0.2,0.2,0.6}
\definecolor{blue}{rgb}{0,0.396,0.741}
\hypersetup{
	colorlinks, 
	bookmarksopen, 
	bookmarksnumbered,
	citecolor=blueRef, 		
	linkcolor=rossoCP3,	
	urlcolor=rossoCP3,			
}

\newskip\humongous \humongous=0pt plus 1000pt minus 1000pt

\newif\ifdtup

\def\oldreffmt#1{\rlap{[#1]} \hbox to 2\parindent{}}

\def\figfmt#1{\rlap{Figure {#1}} \hbox to 1in{}}  
  
\def\etal{\hbox{\it et al.}}  
  
\def\Tr{\mathop{\rm Tr}}

\def\slash#1{#1\!\!\!/\!\,\,} 	
\def\beq{\begin{equation}}  
\def\eeq{\end{equation}}  
\def\bea{\begin{eqnarray}}  
\def\eea{\end{eqnarray}}  
\def\half{\frac{1}{2}}  
  
\def\bq{\begin{quote}}  
\def\eq{\end{quote}}

\def\half{\frac{1}{2}}    

\def \etal {{\it et al.}\ }  
\relax

\newdimen\tdim  
\tdim=\unitlength  
\def\bar{\overline}

\begin{document}

\title{Revisiting Yukawa's Bilocal Field Theory\\
for Composite Scalar Bosons
}

\author{Christopher T. Hill}
\email{hill@fnal.gov}
\affiliation{Particle Theory Department\\
Fermi National Accelerator Laboratory, \\
P. O. Box 500, Batavia, IL 60510, USA}

\begin{abstract}
Yukawa's old bilocal field theory, with modernization in the treatment of ``relative time,''
can describe a relativistic bound state of chiral fermions. This connects to bosonized 
effective chiral Lagrangians and
the Nambu--Jona--Lasinio (NJL) model, providing a description of the internal dynamics
of the bound states.  It features a static internal wave-function, $\phi(\vvr)$, 
in the center-of-mass frame that satisfies a
 Schr\"odinger-Klein-Gordon equation with eigenvalues $m^2$.
We analyze the ``coloron'' model (single perturbative massive gluon exchange)
which yields a UV completion of the NJL model.
This has a BCS-like enhancement of its interaction, $\propto N_c$, the number of colors.
It is {\em classically critical}, with $g_{critical}$ remarkably
close to the NJL quantum critical coupling.  Negative eigenvalues for $m^2$
lead to  spontaneous symmetry breaking, and the Yukawa coupling 
of the bound state to constituent fermions is emergent.
\end{abstract}

\maketitle
 
\date{\today}

\email{hill@fnal.gov}


\section{Introduction }

Many years ago Yukawa proposed a multi local field theory
for the description of relativistic bound states 
\cite{Yukawa}.  
For a composite scalar field, consisting of a pair of constituents,
he introduced a complex bilocal field, $\Phi(x,y)$. This is
factorized,  $\Phi(x,y)\rightarrow \chi(X)\phi(r)$
where $X^\mu=(x^\mu\!+\!y^\mu)/2$ where $r^\mu=(x^\mu\!-\!y^\mu)/2$, and $\chi(X)$ 
 describes the center-of-mass 
of mass motion like any conventional point-like field. Then, $\phi(r)$ 
describes the internal structure of the bound state.
The formalism preserves Lorentz
covariance, though we typically ``gauge fix'' to the center-of mass frame
and Lorentz covariance is not then manifest. Here we must
confront the issue of ``relative time.'' 

Each of the
constituent particles in a relativistic bound state carries its own 
local clock, e.g., $x^0$ and $y^0$. These are in principle independent, so the question
arises, ``how
can a description of a multi-particle bound state be given 
in a quantum theory with a single time variable, $X^0$?''
To this, Yukawa introduced an imaginary ``relative time'' $r^0=(x^0-y^0)/2$, 
but this didn't seem to be effective and
is an element of his construction we will abandon.

 A bilocal field theory 
formalism can be constructed in an action by considering general properties
 of free field bilocal actions.  The bilocal theory is then seen to be directly related to the
local constituent field theory by a introducing a ``world scalar'' constraint into the 
constituent action.  The world scalar is required to match the conserved
currents of the composite theory with those of the constituent theory, and
give canonical normalization to the composite fields.
So, for the composite sector, the constituent action carries a constraint, much
like a gauge fixing condition, directly related to the bilocal field $\Phi(x,y)$
which is a correlation and supplies the constraint.

The removal
of relative time then becomes associated
with the matching of currents and canonical normalization
of the composite fields $\chi$ and $\phi$.
In the center-of-mass frame
the internal wave-function reduces to a static field, $\phi(\vvr)$, where $\vvr=(\vvx-\vvy)/2$.
The approach  leads to a fairly simple
solution to the problem of  relative time,
matching the conclusions one gets from the elegant
Dirac Hamiltonian constraint theory \cite{Dirac}\cite{Reltime}.
The resulting $\phi(\vvr)$ then appears as a  straightforward  result.

We obtain a bilocal bosonic description of a 
composite chiral fermion pair that connects to and provides a UV completion for 
the Nambu--Jona-Lasinio (NJL) model \cite{NJL}, which is recovered in the point-like
limit, $\vvr \rightarrow 0$.
After first considering a bosonic construction,  we apply this to
a theory of chiral fermions with an extended interaction mediated by
a perturbative massive gluon, i.e., the ``coloron model'' 
\cite{Topcolor}\cite{bijnens}\cite{Simmons}\cite{NSD}.   
This leads to an effective (mass)$^2$ Yukawa potential with coupling $g$.
We  form bound states with mass, 
$m^2$, determined as the eigenvalue of a static Schr\"odinger-Klein-Gordon (SKG) equation for
the internal wave-function $\phi(\vvr)$. 

A key result  is a  departure from the usual NJL model, in that
the coloron model has
a nontrivial {\em classical critical behavior}, $g>g_c$, 
leading to a bound state with a negative $m^2$.  The classical interaction
is analogous to the Fr\"ohlich Hamiltonian interaction in a superconductor and
has a BCS-like enhancement of the coupling by a factor of $N_c$ (number of colors) \cite{Cooper}.
Remarkably we find the classical $g_c$ is numerically close to the NJL critical coupling constant
which arises in fermion loops. 

Moreover, the scalar bound state will have an effective Yukawa coupling to its constituent fermions,
distinct from $g$, 
that is emergent in the theory.  In the point-like limit this matches the NJL coupling when near
criticality.  However, in general this depends in detail upon the internal wave-function
$\phi(\vvr)$ and potential.  In the point-like limit this is determined
by $\phi(0)$ and we recover the NJL model.  
However, if we are far from the point-like limit in an extended wave-function
$\phi(\vvr)$ might suppress the emergent
Yukawa coupling, even though the coloron coupling $g$ is large. 

The description of a relativistic bound state in the rest frame is therefore
similar to the eigenvalue problem of the non-relativistic Schr\"odinger equation
and some intuition carries over.  However,
the eigenvalue of the static Schr\"odinger-Klein-Gordon (SKG) 
equation, is $m^2$, rather than energy. Hence, a 
bound state with positive $m^2$  is a resonance that can decay to its
constituents and has a Lorentz line-shape in $m^2$ (we give an example in
Appendix \ref{barrier}), and thus has a large
distance radiative component in its solution
that represents incoming and outgoing open scattering states. 

If the eigenvalue for $m^2$ is negative, or tachyonic, then contrary to
the non-relativistic case the bound state represents a
chiral vacuum instability. This then requires consideration of a
quartic interaction of the composite field, $\sim \lambda \Phi^4$ which is
expected to be generated by the underlying theory (which we only treat phenomenologically
in the present paper).  
In the broken symmetry phase the composite field $\Phi(x,y)$ acquires a vacuum expectation value (VEV),
$\langle\Phi\rangle=v$.  It is {\em a priori} less
clear what happens to the internal wave-function $\phi(r)$ which then becomes the solution
of a nonlinear SKG equation
i.e.,  does the internal wave-function, $\phi(\vvr)$, significantly deform 
and acquire a sympathetic VEV as well,
or does it remain localized?
In the perturbative ($\lambda$) solution, 
in the broken phase $\phi(\vvr)$ remains localized 
and the Nambu-Goldstone modes and Brout-Englert-Higgs (BEH)
boson retain the common localized solution for their internal wave-functions.

\section{ Constructing a Bilocal Composite Theory }

\subsection{Brief Review of the NJL Model}

The Nambu--Jona-Lasinio model (NJL) \cite{NJL} is the simplest field
theory of a composite scalar boson,
consisting of a pair of chiral fermions. A bound state emerges 
from an assumed point-like 4-fermion interaction and
is described by  local effective field, $\Phi(x)$. The effective field arises as an auxiliary field
from the factorization of the 4-fermion interaction.
In the usual formulation of the NJL model, chiral fermions induce loop effects
in a leading large $N_c$ limit which, through the renormalization group (RG),
lead to the interesting dynamical 
phenomena at low energies. 
We give a lightning review of this presently.

We assume chiral fermions, each with $N_c$ ``colors'' labeled by $(a,b,...)$.
A non-confining, point-like  chiral invariant  $U(1)_{L}\times U(1)_{R}$
interaction, then takes the form:
\smbbo
\label{NJL1}
S_{NJL}
=\!\int\! d^4x \;\bl i\bar{\psi}^a_L(x)\slash{\partial}\psi_{aL}(x)
+ i\bar{\psi}^a_R(x)\slash{\partial}\psi_{aR}(x)
\nonumbo
\qquad \qquad
+\;
\frac{g^2}{M_0^2}
\bar{\psi}^a_L(x)\psi_{aR}(x)\;\bar{\psi}^b_R(x)\psi_{bL}(x)
\br.
\smebo
This can be readily generalized to a $G_L\times G_R$ chiral symmetry.
We then factorize eq.(\ref{NJL1}) 
by introducing the local auxiliary
field $\Phi(x)$ and write for the interaction:
\smbbo
\label{NJL2}
\backo
\int \!d^4x \bl g\bar{\psi}^a_L(x)\psi_{aR}(x)\Phi(x)+h.c. - M_0^2\Phi^\dagger(x) \Phi(x) \br.
\smebo 
We view eq.(\ref{NJL2}) as the action defined at the high scale $\mu \sim M$.
We then,  following \cite{Wilson}, integrate out the fermions to obtain the effective action for the composite 
field $\Phi$ at a lower scale $\mu<\!\!<M$. 

{
\begin{figure}
	\centering
	\includegraphics[width=0.4\textwidth]{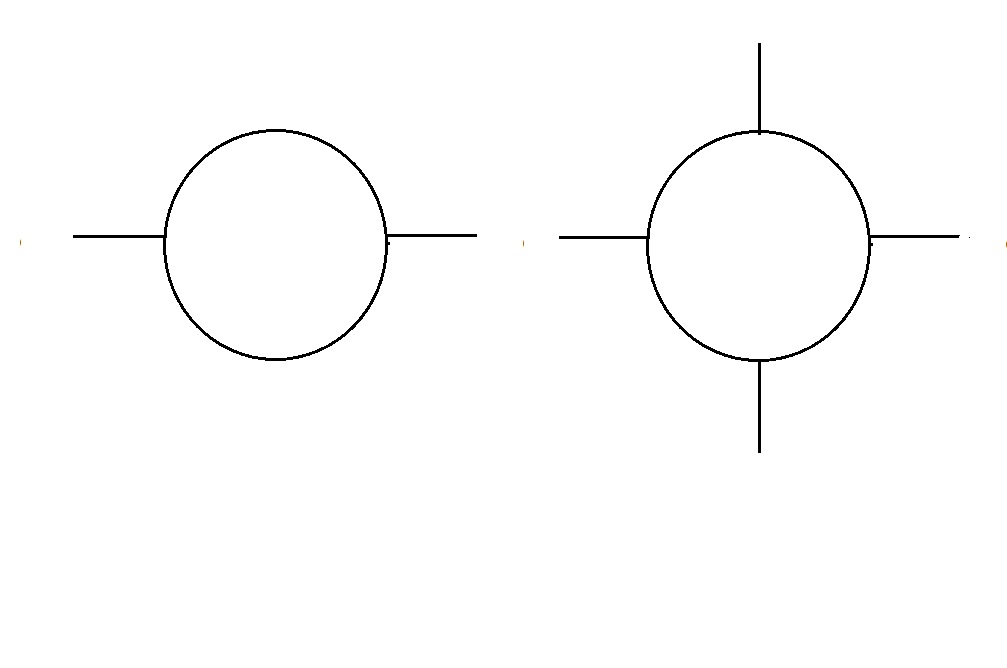}
	\vspace{-0.6in}
	\caption{ Diagrams contributing to the point-like NJL model effective
	Lagrangian, eqs.(\ref{NJL2},\ref{NJL3}). External lines are $\Phi$ and internal
	lines are fermions $\psi$.}
	\label{figvar}
\end{figure}
}

The calculation in the large-$N_c$ limit and full RG
is discussed in detail in \cite{BHL,CTH}.  
The leading $N_c$ fermion loop yields the result:
\smbbo
\label{NJL4}
\backo\!\!\!
L_{M}\rightarrow  L_{\mu }= g[\bar{\psi}_{R}\psi_L ]\Phi + h.c 
+Z\partial_\mu \Phi^\dagger\partial^\mu \Phi
\nonumbo
\qquad\qquad 
-m^2\Phi ^{\dagger }\Phi - \frac{\lambda }{2}(\Phi^\dagger \Phi)^2,
\smebo
where,
\smbbo
\label{NJL3}
\backo
m^2 = M_0^{2}\!-\!\frac{N_{c}g^{2}}{8\pi ^{2}} ( M_0^{2}-\mu^2),
\nonumbo
\backo
Z=\frac{N_{c}g^{2}}{16\pi ^{2}}\ln(M_0/\mu), \;\;\;
\lambda=\frac{N_{c}g^{4}}{8\pi ^{2}}\ln( M_0/\mu).
 \smebo
Here $M_0^2$ is the UV loop momentum 
cut-off, and we include the induced kinetic and quartic interaction terms.
The one-loop result can be improved by using the full RG \cite{BHL,CTH}.
Hence the NJL model is driven by fermion loops, which are $\propto \hbar$ intrinsically
quantum effects.

Note the behavior of the composite scalar boson mass, 
$m^2$, of eq.(\ref{NJL3}) in the UV.
The $ -{N_{c}g^{2}M_0^{2}}/{8\pi ^{2}}$  term arises from the negative
quadratic divergence in the loop involving the
pair $\left( {\psi }_{R},\psi _{L}\right) $ of Fig.(1), with
 UV cut-off  $M_0^{2}$. 
Therefore, the NJL model has a critical
value of its coupling defined by the cancellation of the large $M_0^2$ terms
for $\mu^2=0$,
\smbbo
g_{c0}^{2}=\frac{8\pi ^{2}}{N_{c}}+{\cal{O}}\bl\!\frac{\mu^2}{M^2}\!\br.
\smebo
Note that $\mu$ is the running RG mass, and comes from the lower limit
of the loop integrals and breaks scale invariance and can in principle be small, 
$\mu^2< \!\!<m^2$. 
Hence $m^2$ is always zero and the theory scale
invariant at the critical value, corresponding to a second order phase transition. 

For  super-critical coupling, $g>g'_c$ 
we see that $m^2<0$ and there will be a vacuum instability.
The effective action, with a $\lambda |\Phi|^4$ term, is then the usual sombrero potential,
The chiral symmetry is spontaneously broken, the 
chiral fermions acquire mass, and the theory generates 
Nambu-Goldstone bosons. Fine tuning of $g^2 \approx g_c^2$ is possible 
if we want a theory with a hierarchy, $|m^2|<\!\!<M_0^2$.

\subsection{Construction of Bilocal Composites in a Local Scalar Field Theory  $\label{bloc}$ } 

Presently we will obtain a description of bound states by bilocal fields
in a Lorentz invariant model, consisting of a point-like complex scalar field and an interaction 
mediated by a point-like real field (in section {\ref{3}} we extend this to a chiral fermions
interacting by a massive gauge field, analogous to a heavy gluon, aka ``coloron''; in 
Appendix ($\ref{SumNote}$) we give a summary of notation and formulas).
We 
will see that we are led to a Schr\"odinger-Klein-Gordon equation for a static 
internal bound state wave-function
in the center of mass frame.
Our present treatment will be semi-classical.

Consider local scalar fields $\varphi(x)$ (complex) and $A(x)$ (real) and action:
 \smbbo 
 \label{start}
\backo\!\!\!
S=\int_x\bl |\partial\varphi|^2 \!+\! \half (\partial A)^2\!-\!\half M^2A^2\!-\!gM|\varphi|^2A
\!-\!\frac{\lambda}{2}|\varphi|^4\br,
 \smebo
 where we abbreviate, $|\partial\varphi|^2=\partial_\mu\varphi^\dagger\partial^\mu\varphi$
 and $(\partial A)^2= \partial_\mu A\partial^\mu A$.
Here $g$ is dimensionless and we refer all mass scales to the single scale $M$. 
We will discuss the quartic term separately below, and presently set it aside, $\lambda=0$.

If we integrate out $A$ we obtain an effective, attractive, bilocal potential interaction term
at leading order in $g^2$, 
 \smbbo
\label{act00}
\backo\backo
S=\!\!\int_x\! |\partial\varphi|^2 
+\frac{{g^2M^2}}{2}\!\!\!\int_{xy}\!\!\!\! \varphi^\dagger\!(y)\varphi(y)D_F(y\!-\!x)
\varphi^\dagger\!(x)\varphi(x),
 \smebo
where the two-point function is given by $(i)\times$ the Feynman propagator, 
 \smbbo
 \label{DF}
D_F(x-y)=-\!\!\int \frac{e^{iq_\mu(x^\mu-y^\mu)}}{(q^2-M^2)} \frac{d^4q}{(2\pi)^4}.
 \smebo
The equation of motion of $\varphi$ is therefore,
 \smbbo
\label{acteqn1}
\backo\backo
\partial^2\varphi(x) -g^2M^2\!\!\int_{y}\!\! \varphi^\dagger(y)\varphi(x)D_F(x-y)\varphi(y)=0,
 \smebo
 (note we have transposed $\varphi^\dagger(y)$ and $\varphi(y)$ under the integral).
In the action eq.(\ref{act00}), the kinetic term is still
local while the interaction is bilocal, and the theory is still classical in
that this only involved a tree diagram that is  ${\cal{O}}(\hbar^{0})$.

We now define a bilocal field of mass dimension $d=1$,
 \smbbo 
\label{biloc}
\Phi(x,y)=M^{-1}\varphi(x)\varphi(y) .
 \smebo
The free particle states described by the bilocal field trivially satisfy
an equation of motion,
\smbbo
 \label{eq1}
 \partial_x^2\Phi(x,y)=0 ,
 \smebo
 and this is generated by a bilocal action,
 \smbbo
\label{act0}
\backo
S=M^4\!\!\int_{xy}\!\! Z|\partial_x\Phi(x,y)|^2,
\smebo
where we will specify the normalization, $Z$ and scale $M$ subsequently  (we discuss the general properties
of the bilocal fields and actions in Appendices B.2 and B.3).
With the bilocal field the interaction of eq.(\ref{act00}) becomes,
 \smbbo
 \frac{{g^2M^4}}{2}\!\!\!\int_{xy}\!\!\!\! \Phi^\dagger(x,y)D_F(x\!-\!y) \Phi(x,y).
 \smebo

We can therefore postulate a bilocalized action as a free particle
part plus the interaction,
\smbbo
\label{act0}
\backo
S=\!\!\int_{xy}\!\!\bl ZM^4|\partial_x\Phi(x,y)|^2 \!
+\!\half{g^2M^4}\Phi^\dagger(x,y)D_F(x-y)\Phi(x,y)\br.
\nonumbo
\smebo
In the limit $g=0$ the field $\Phi(x,y)$ and the action faithfully represents
two particle kinematics, and we have the equation of motion,
\smbbo
\label{15}
\backo
0=Z\partial^2_x\Phi(x,y) 
-\half{g^2}D_F(x-y)\Phi(x,y).
\smebo
We see that a $U(1)$ conserved Noether current is
generated by $\Phi(x,y)\rightarrow e^{i\theta(x)}\Phi(x,y)$,
\smbbo
J_{\Phi\mu}(x)= 
iZ\int d^4y\bl \Phi^\dagger (x,y) \frac{\stackrel{\leftrightarrow}{\partial}}{\partial x^\mu}\Phi(x,y)\br,
\smebo
where $A\stackrel{\leftrightarrow}{\partial}B=
A{\partial}B-({\partial}A)B$. This must match the conserved $U(1)$  current in
the constituent theory, 
\smbbo
J_{\varphi \mu}(x)= 
i\varphi^\dagger (x) \frac{\stackrel{\leftrightarrow}{\partial}}{\partial x^\mu}\varphi(x).
\smebo
Substituting  eq.(\ref{biloc}) into $J_{\Phi\mu}(x)$ we see that the matching requires,
 \smbbo
\label{match}
J_{\Phi\mu}(x)=J_{\varphi \mu}(x)\; ZM^{2} \!\!\int \!d^4y |\varphi(y)|^2,
 \smebo
hence,
 \smbbo
\label{norm000}
1=
ZM^{2}\int \!d^4y \;|\varphi(y)|^2.
 \smebo  
 This is a required constraint for the bound state sector of the theory.\footnote{This 
 normalization reflects the asymmetric kinetic term we have chosen, $\sim\partial_x^2$. 
 This breaks the symmetry $\Phi(x,y) =\Phi(y,x)$ and effectively treats one of the particles
 as having zero charge.  
 We will discuss currents 
 in greater deal elsewhere. }
  Note that the square of the constraint is the 4-normalization of $\Phi$,
  \smbbo
\label{normphi}
\backo
1=
Z^2M^{4}\int \!d^4y\; d^4y \;|\varphi(y)|^2|\varphi(x)|^2
\nonumbo
=
Z^2M^6\int \!d^4y\;d^4y \;|\Phi(x,y)|^2.
 \smebo
 This implies that the presence of a correlation in the two particle sector, $\Phi(x,y)$
 acts as a constraint on the single particle action in that sector.
 We can now see how the underlying $\varphi$  action of eq.(\ref{act00})
 leads to the $\Phi$ action by inserting the constraint of eq.(\ref{norm000}) 
 onto the kinetic term of eq.(\ref{act00})
 and rearranging to obtain,
 \smbbo
\label{act0001}
\backo
S=\!\!\int_{xy}\!\! \bl ZM^{2}|\varphi(y)\partial_x\varphi(x)|^2 
\nonumbo
\qquad \!\!\!
+{\frac{g^2M^2}{2}}\varphi^\dagger(y)\varphi(x)D_F(x-y)\varphi^\dagger(x)\varphi(y)\br,
 \smebo
 and $S$ remains  dimensionless. With 
 the bilocal field of eq.(\ref{biloc})
 the bilocalized action eq.(\ref{act0001}) becomes eq.(\ref{act0}).

\vspace{0.2in}
Following Yukawa, we go to barycentric coordinates  $(X,r)$,
 \smbbo
\label{barycentric1}
\backo
X=\half(x+y),\qquad r=\half(x-y).
 \smebo
where $r^\mu=(r^0,\vvr)$, where $\vvr$ is the radius and $r^0$ is the relative time.\footnote{Yukawa
preferred to write
things in terms of $\rho=2r$ 
which has the advantage of a unit Jacobian, $\int d^4x d^4y
= J\int d^4X d^4\rho$ with $J=1$. We find that the radius, $r$, is more convenient in loop
calculations and derivatives are symmetrical, $\partial_{X,r}=(\partial_x \pm \partial_y)/2$
vs. $=\half\partial_X +\partial_\rho $, 
but requires the Jacobian. See Appendix ($\ref{SumNote}$) for a summary of notation.
} 
Hence we write,
 \smbbo
\backo
\Phi(x,y)=\Phi(X\!+\!r,X\!-\!r)\equiv
\Phi(X,r).
 \smebo
 Let  $S=S_K+S_P$ 
and can then rewrite 
the kinetic term, $S_K$,
using the derivative $\partial_x=\half(\partial_X\!+\!\partial_r)$, 
\smbbo
\label{act000}
\backo
S_K=
\frac{JM^4}{4} \!\! \int_{Xr}\!\! Z|(\partial_X\!+\!\partial_r)\Phi(X,r)|^2 \!
\nonumbo  \backo\!\!\!\!\!
=\frac{JM^4}{4} \!\!\int_{Xr}\!\bl Z|\partial_X\Phi|^2\!+\! Z|\partial_r\Phi|^2 
\!+\! Z(\partial_X\Phi^\dagger\partial_r\Phi+h.c.)\br.
\smebo 
Note the Jacobian $J=16$,
\smbbo
J^{-1}=\left| \frac{\partial (X, r)}{\partial(x,y)} \right|=\bl \frac{1}{2}\br^4.
\smebo
Likewise, the potential term is,
 \smbbo
S_P=
\frac{JM^4}{2}\!\int_{Xr}\! {g^2}D_F(2r)|\Phi(X,r)|^2.
 \smebo
 
We will treat the latter term in eq.(\ref{act000}),
$
Z(\partial_X\Phi^\dagger\partial_r\Phi+h.c.),
$
as a constraint, 
with its contribution to the equation of motion,
\smbbo
\frac{\partial}{\partial X^\mu}\frac{\partial}{\partial r_\mu}\Phi=0.
\smebo
  We can redefine this term in the action as a Lagrange multiplier while
  preserving Lorentz invariance,
\smbbo
\label{LM}
\backo
\rightarrow \int_{Xr}\eta \bl
\frac{\partial\Phi^\dagger}{\partial X^\mu}\frac{\partial\Phi}{\partial r_\mu}+h.c.\br^2\;\;
 \makebox{hence,} \;\; \delta S/\delta \eta = 0,
 \smebo
which also enforces the constraint on a path integral in analogy to gauge fixing.
In the following we assume the constraint is present in the total action though not
written explicitly.
 
We therefore have the bilocal action, 
\smbbo
\label{23}
\backo\backo
S=
\frac{JM^4}{4}\!
\!\int_{Xr}\!\!\!\!\!(
Z|\partial_X\Phi|^2+ Z|\partial_r\Phi|^2 \!
+2g^2 D_F(2r)|\Phi(X,r)|^2).
\smebo
Following Yukawa, we assume we can factorize $\Phi$,
 \smbbo
\label{factor000}
\sqrt{J/4}\; \Phi(X,r)= \chi(X)\phi(r).
 \smebo
${\phi(r)}$ is the internal wave-function which we define to be dimensionless, $d=0$, 
while $\chi$ is an ordinary local field 
with mass dimension $d=1$.
$\chi(X)$ determines the center-of-mass motion of the composite state.
The action for the factorized field takes the form, 
\smbbo
\backo\backo
S=
{M^4}\!
\!\int_{Xr}\!\bl\!
Z|\partial_X\chi|^2|\phi^2|
\nonumbo
 + |\chi|^2( Z|\partial_r\phi|^2 \!
+2g^2 D_F(2r)|\phi(r)|^2\! ) \br.
\smebo
Matching the $U(1)$ current generated by 
$\chi\rightarrow e^{i\theta(X)}\chi$
(or to have a canonical normalization of $\chi(X)$), we see that the normalization of the
world-scalar 4-integral is,\footnote{ The current is generated by
$e^{i\theta(x)}\Phi=e^{i\theta((X+r)/2)}\Phi$. The matching is then exact for the zero component (charge)
an implies eq.(\ref{chiact2}).
There is a correction to the spatial component 
$\propto \phi^\dagger\stackrel{\leftrightarrow}{\partial_{\vvr}} \phi$. The currents will be
discussed in detail elsewhere.}
\smbbo
\label{chiact2}
1 =ZM^4 \! \int\! d^4r\; |\phi(r)|^2.
 \smebo

We can then represent $S$ in terms of two ``nested'' actions. For the field $\chi$,
 \smbbo
\label{chiact}
\backo
S=
\!\int_{X}\!\bl |\partial_X\chi|^2-m^2|\chi|^2\br
\;\;\;\makebox{where,}\;\;\; m^2=- S_\phi,
 \smebo
and  $S_\phi$ is an action for the internal wave-function,
\smbbo
\label{act0p}
\backo\!\!\!
S_\phi=M^4\!\!\int_{r^0,\vvr}\!\bl\! Z|\partial_r\phi|^2\!
+{2g^2 }D_F(2r)|\phi|^2\!\br.
\smebo
Eq.(\ref{chiact}) then implies,
 \smbbo 
\partial^2_X\chi=-m^2\chi \;\;\;\makebox{hence},\;\;\chi\sim \exp(iP_\mu X^\mu).
 \smebo
$\chi(X)$ has free  plane wave solutions with $P^2=m^2$.

In the center of mass frame of the bound state we can choose $\chi$ to
have 4-momentum $P_\mu=(m,0,0,0)$
where we then have,
 \smbbo
\label{factorize}
\Phi(X,r)=\chi(X)\phi(r)\propto \exp(i m X^0)\phi(r).
 \smebo
$\phi(r)$ must then satisfy the Lagrange multiplier,
constraint 
\smbbo
P^\mu \frac{\partial}{\partial r^\mu} \phi(r) =0,
\smebo
and therefore becomes a {\em static function} of 
$r^\mu=(0,\vvr)$. 

While we have specified $Z$ in eq.(\ref{chiact2}), we still have 
the option of normalizing the internal
wave-function
$\phi(\vvr)$. This can be conveniently 
normalized in the center of mass frame as,
 \smbbo
\label{norm01}
\backo\!\!\!\!\! M^3
\!\int \!\! d^3r\; |\phi(\vvr)|^2=1.
\smebo
Note that in eq.(\ref{norm01}) we have implicitly
defined the static internal wave-function $\phi(\vvr)$ to be 
dimensionless, $d=0$.

We see that the relative time now emerges in the 4-integral over $|\phi(r)|^2$
of eq.(\ref{chiact2}) together with eq.(\ref{norm01}),
 \smbbo
\label{norm0000}
\backo\backo
1=ZM^4\!\!\int\!\! d^4r|\phi(r)|^2=ZM^4\!\!\int \!\!dr^0 \!\!\!\int\!\! d^3r |\phi(\vvr)|^2=ZMT,
 \smebo
 where $\int dr^0= \int dr^\mu P_\mu/m \equiv T$.
Then from eq.(\ref{norm0000}) we have,
\smbbo
\label{norm02}
TZ=M^{-1}.
 \smebo
With static $\phi(r)\rightarrow \phi(\vvr)$ the internal  action of eq.(\ref{act0p}) becomes,
\smbbo
\label{SKG0}
\backo
S_\phi=M^4\!\!\int_{r^0,\vvr}\!\bl\!-Z|\partial_{\vvr}\phi(\vvr)|^2\!
+{2g^2 }D_F(2r)|\phi(\vvr)|^2\!\br,
\smebo
where $|\partial_{\vvr}\phi(\vvr)|^2=\partial_{\vvr}\phi^\dagger\cdot\partial_{\vvr}\phi$.
Note  $\partial_{\vvr}\phi$ is spacelike,
and the arguments of the constrained $\phi(\vvr)$ are now 3-vectors,
however $D_F(2r^\mu)$ still depends upon the 4-vector $r^\mu$.

There remains the integral over relative time $r^0$ in the action.
For the potential, we have by residues,
\smbbo
\label{42}
\backo\!\!\!\!\!\!\!  -V(r)=2\!\!\int \!\!dr^0 D_F(2r)
 =\!\int \!\!\frac{e^{2i\vvq\cdot\vvr}}{{\vvq}^2+M^2} \frac{d^3q}{(2\pi)^3}
 =\!\frac{e^{-2M|\vvr|}}{8\pi |\vvr|},
\smebo 
and the potential term becomes the static Yukawa potential,
\smbbo
\backo
S_P=-M^3\!\!\int_{\vvr}\!
{g^2 }MV(\vvr)|\phi(\vvr)|^2,\qquad
V(\vvr)=-\frac{e^{-2M|\vvr|}}{8\pi|\vvr|}.
\smebo
The $\phi(\vvr)$ kinetic term in eq.(\ref{SKG0}) becomes,
\smbbo
\backo
S_K=-M^4\!\!\int_{r^0,\vvr}\!\!Z|\partial_{\vvr}\phi(\vvr)|^2\!
=-M^4ZT\!\! \int_{\vvr}\! |\partial_{\vvr}\phi(\vvr)|^2\!
\nonumbo
=-M^3\!\!\int_{\vvr}\! |\partial_{\vvr}\phi(\vvr)|^2\!,
\smebo
where we use eq.(\ref{norm02}).
The action $S_\phi$ thus becomes, 
\smbbo
\label{act0ppp}
\backo\!\!\!
m^2=
-S_\phi= M^{3}\int_{\vvr} \bl |\partial_{\vvr}\phi|^2
+{g}^2MV(|\vvr|)|\phi|^2 \br.
\smebo
Note $S_\phi$ has dimension $d=2$, as it must for $m^2$.
We thus see, as previously mentioned, that the combination $ZT$ occurs in the theory, and 
the relative time has disappeared into normalization constraints,
eqs.(\ref{chiact2}, \ref{norm01}).

The extremalization of $S_\phi$ leads to the  Schr\"odinger-Klein-Gordon (SKG) equation,
\smbbo
\label{acteqn1}
\backo
-\nabla_{\vvr}^2\phi(r) -{g^2M}\frac{e^{-2M|\vvr|}}{8\pi|\vvr|}\phi(r)= m^2{\phi}(r),
\smebo
where, for spherical symmetry in a ground state,
\smbbo
\nabla_r^2=\partial_r^2+\frac{2}{r}\partial_r.
\smebo
We see that the induced mass$^2$ of the bound state, $m^2$,
is the eigenvalue of the SKG equation.
We can compare this to a non-relativistic Schr\"odinger equation (NRSE),
 \smbbo
\label{acteqn1}
\backo
-\frac{1}{2M}\nabla_r^2{\phi}(r) - {g^2}\frac{e^{-2Mr}}{16\pi r}{\phi}(r)= E{\phi}(r).
 \smebo
 In the next section we will obtain a similar results for
 a bound state of chiral fermions and use the known results for the Yukawa potential
 in the NRSE to obtain the critical coupling.
 The negative eigenvalue of $E$ in the NRSE, which signals binding, presently implies
 a vacuum instability.
 
Integrating by parts we then have from eq.(\ref{act0ppp}),
\smbbo
\label{act0pppp}
\backo\!\!\!
m^2 =M^{3}\int_{\vvr} \bl \phi^\dagger (-\nabla^2_{\vvr}\phi
+{g^2M}V(r)\phi(\vvr))\br.
\smebo
Note consistency, using eq.(\ref{acteqn1}),
and the normalization of the dimensionless field $\phi$
 of eq.(\ref{norm01}).

More generally, 
by promoting $\chi$ to a $(1+3)$ time dependent field while maintaining a static $\phi$
we have the full joint action:
 \smbbo
\backo\;
S=\!
M^3\!\!\int_{X\vvr} \!\bl \!|\phi|^2\!\left|\frac{\partial\chi}{\partial X}\right|^2\!
\!\!-\!|\chi|^2\bl |\partial_{\vvr}\phi|^2\!
+\!g^2M V(\vvr)|\phi|^2\br\br .
\nonumbo
 \smebo

In summary, we have constructed, by ``bilocalization''
of a local  field theory, a bilocal field description
$\Phi(x,y)$ for the dynamics of binding
a pair of particles.  The dynamics
implies that, in barycentric coordinates, $\Phi(x,y)\sim \Phi(X,r)\sim \chi(X)\phi(\vvr)$,
where the internal wave-function, $\phi(\vvr)$, is a static function of $\vvr$
and satisfies an SKG equation with eigenvalue $m^2$, which determines the squared-mass of
a bound state.  This illustrates the removal of relative time in an action formalism, which
is usually framed in the context of Dirac Hamiltonian constraints \cite{Dirac}.

\subsection{ Simplified Normalization}

The normalization system we have thus far used is awkward.  We can facilitate this
by defining a new integral over the internal wave function 3-space $\vvr$:
 \smbbo
\label{compact0}
\int_{r}'\equiv   \int \frac{d^3r}{U}\;\;\makebox{where}\;\; U= M^{-3}.
 \smebo
We then have the key elements of the theory in this notation:
\smbbo
\label{compact1}
S=
\!\int \!\!d^4X\bl |\partial_X\chi|^2-m^2|\chi|^2\br
\nonumbo
1=\int'_{r} |\phi(\vvr)|^2= \int \!\frac{d^3r}{U} |\phi|^2
\nonumbo
m^2=-S_\phi
\nonumbo
S_\phi= \int'_{r} \bl -|\partial_{\vvr}\phi|^2
-{g^2M}V(r)|\phi|^2\br
\nonumbo
 V(r)
 =-\frac{e^{-2M|\vvr|}}{8\pi |\vvr|}.
\smebo
Our notation is summarized in Appendix \ref{SumNote}.

\section{The Coloron Model $\label{3}$}

\subsection{Boundstate and $N_c$-Enhanced Coupling $\label{BCS}$ }

  The point-like NJL model can be viewed as the limit of a physical theory
with a bilocal interaction.
An example  that
 motivates the origin of the NJL
interaction is an analogue of QCD, with a massive and perturbatively coupled gluon.
We call this  a ``coloron model,'' and it has been extensively deployed
to describe chiral constituent quark models \cite{bijnens}\cite{bardeenhill}, and
to discuss the possibility of the BEH boson composed of top quarks \cite{Topcolor},
and as a generic model for experimental search strategies \cite{ NSD}\cite{Simmons}.

Consider a nonconfining $SU(N_c)$ gauge
theory,  broken to global $SU(N_c)$, where
the coloron gauge fields, $A_\mu^A$, acquire mass, $M$,
and have a fixed coupling constant $g$. 
We assume chiral fermions, each with $N_c$ ``colors'' labeled by $(a,b,...)$
with the local Dirac action,
 \smbbo
 \label{fermikinetic}
\backo
S_F= \int_x\bl i\bar{\psi}^a_L(x)\slash{D}\psi_{aL}(x)+ i\bar{\psi}^a_R(x)\slash{D}\psi_{aR}(x)\br,
 \smebo
where the covariant derivative is,
\smbbo
D_\mu=\partial_\mu - igA^A_\mu(x)T^A,
\smebo
and $T^A$ are the adjoint representation generators of $SU(N_c)$.
We assume the colorons have a common mass $M$.

The single coloron exchange interaction   
 then takes a bilocal current-current form:
\smbbo
\label{TC0}
 \backo\;
S_C=-g^2\!\!\!\int_{xy} \!\!
 \bar{\psi}_{L}(x) \gamma_\mu T^A \psi_L(x)\;D^{\mu\nu}(x-y)\;
\bar{\psi}_{R}(y) \gamma_\nu T^A \psi_{R}(y),
\nonumbo
\smebo
where $T^A$ are generators of $SU(N_c)$. 
The coloron propagator  in a Feynman gauge yields:
\bea
\label{propagator}
D_{\mu\nu}(x-y)=\int\frac{-ig_{\mu\nu}}{q^2-M^2}e^{iq(x-y)}\frac{d^4q}{(2\pi)^4}.
\eea
A Fierz rearrangement of the interaction to leading order in $1/N_c$ 
leads to an attractive potential \cite{Topcolor}:
\smbbo
\label{coloronexchange}
\backo
S_C=g^2\!\!\int_{xy}\!\! \; \bar{\psi}^a_L(x)\psi_{aR}(y)\; D_F(x-y)\;\bar{\psi}^b_R(y)\psi_{bL}(x),
\smebo
where $D_F$ is defined in eq.(\ref{DF}).
Note that if we suppress the $q^2$ term in the denominator
of eq.(\ref{propagator}), 
 \smbbo
\label{propagator2}
{D}_F(x-y)\rightarrow  \frac{1}{M^2}\delta^4(x-y),
 \smebo
and  we  immediately recover the point-like NJL model interaction.

Consider spin-$0$ fermion pairs of a given color $[\bar{a}b]$
$ \bar{\psi}^a_R(x)\psi_{bL}(y)\!$.
We 
will have free fermionic scattering states, $\sim\; :\! \bar{\psi}^a_R(x)\psi_{bL}(y)\! :$,
coexisting in the action with bound states $\Phi(x,y)$,
 \smbbo
\label{act2}
\backo\!\!\! \bar{\psi}^a_R(x)\psi_{bL}(y)\!
\rightarrow \;:\! \bar{\psi}^a_R(x)\psi_{bL}(y) \!:+  \;\Phi^a_{\;b}(x,y).\;\;
 \smebo 
 This is analogous to the shift done to introduce the auxilliary field
 in the NJL model in eq.(\ref{NJL2}). The main differences with the NJL model are that presently
 $\Phi^a_{\;b}(x,y)$ is bilocal and it has a bare two body kinetic term.
 The normal ordering $:...:$ signifies that we have subtracted
 the bound state from the product.

We see that
$\Phi^a_b(X,r)$ is an $N_c\times N_c$ complex matrix that transforms as a product
of $SU(N_c)$ representations,  $\bar{N}_c\times  {N}_c$,
and therefore decomposes into a singlet plus an adjoint representation of $SU(N_c)$.
We write $\Phi^a_b$
it as a matrix  $\widetilde\Phi$  by introducing the $N_c^2-1$ adjoint matrices,
${T}^A$, where $\Tr ({T}^A{T}^B)=\half\delta^{AB}$.
The unit matrix is ${T}^0\equiv\;$diag$(1,1,1,..)/\sqrt{2N_c}$, 
and $\Tr({T}^0{})^2 = 1/2$, hence we have,
\smbbo
\widetilde{\Phi} =\sqrt{2}({T}^0\Phi^0 + \sum_{A}{T}^A\Phi^A).
\smebo
The $\sqrt{2}$ is present because $\Phi^0$ and $\Phi^A$ form complex representations  
since they also represent the $U(1)_L\times U(1)_R$
chiral symmetry.

For the bilocal fields, 
we have a  bosonic kinetic  term  with the constraint,
\smbbo
\label{act0f}
\backo\backo
S_K=\frac{{J}ZM^4}{2} \!\int_{Xr}\!\! \Tr\bl  |\partial_X\widetilde\Phi|^2
+|\partial_r\widetilde\Phi|^2
\!+\! \eta|\partial_X\widetilde\Phi{}^\dagger\partial_r\widetilde\Phi|^2
\br.
\smebo
Note the numerical factor differs
from the scalar case
by treating $(x,y)$ symmetrically as in eq.(\ref{a1}).
For the singlet representations this takes the form,
\smbbo
\label{act1}
\backo\backo
S_K= \frac{{J}ZM^4}{2}\!\!\int_{Xr}\! \bl|\partial_X\Phi^0|^2+|\partial_r\Phi^0|^2
\!+\! \eta|\partial_X\Phi^0{}^\dagger\partial_r\Phi^0|^2
\br .
\smebo
We assume the constraint in the barycentric frame, and integrate out
relative time with $ZMT=1$,
\smbbo
\label{act1}
\backo\backo
S_K= (J/2)\!\!\int_{X}\int'_{\vvr}\! \bl|\partial_X\Phi^0(X,\vvr)|^2-|\partial_{\vvr}\Phi^0(X,\vvr)|^2
\br,
\smebo
(where $\int'_{\vvr} =M^3\int d^3r$).

The full interaction of eq.(\ref{coloronexchange}) thus becomes,
\smbbo
\label{interaction0000}
\backo
\;S_C
\rightarrow
g^2\!\!\int_{xy}\!\! :\!\bar{\psi}^a_L(x)\psi_{aR}(y)\!: D_F(x-y) :\!\bar{\psi}^b_R(y)\psi_{bL}(x)\!:
\nonumbo 
+\;g^2JM^2\sqrt{N_c}\!\!\int_{X,r}\!\!\! :\!\bar{\psi}^a_L(X\!-\!r)\psi_{aR}(X\!+\!r)\!: D_F(2r)\;\Phi^0{+h.c.}
\nonumbo
+\;g^2JM^4N_c\!\!\int_{X,r}\!\!\!\! \; \Phi^0{}^\dagger(X,r)\; D_F(2r)\;\Phi^0(X,r),
\smebo
where,
 \smbbo
{D}_F(2r)= -\!\int\frac{1}{(q^2-M^2)}e^{2iq_\mu r^\mu}\frac{d^4q}{(2\pi)^4}.
 \smebo
The leading term $S_C$ of eq.(\ref{interaction00}) is just a free 4-fermion
scattering state interaction 
and has the structure of the NJL interaction in the limit of eq.(\ref{propagator2}).
This identifies $g^2$ as the NJL coupling constant.
This is best treated separately by the local interaction of eq.(\ref{coloronexchange}). 
We therefore omit this term in
the discussion of the bound states.

The second term $\sim \Tr(\psi^\dagger\psi)\Phi^0+h.c.$ in eq.(\ref{interaction00}) 
determines the Yukawa interaction between
the bound state $\Phi^0$ and the free fermion scattering states.
We will treat this below.

Note that the third term in eq.(\ref{interaction0000})  is the binding interaction and it
involves only the singlet, $\Tr\tilde\Phi = \sqrt{N_c}\Phi^0$. 
The adjoint representation 
 $\Phi^A$ are decoupled from the interaction
 and remain as unbound, two body massless scattering states. 
Moreover, the singlet field $\Phi^0$ therefore has an enhanced
 interaction by a factor of $N_c$.
This is analogous to a BCS superconductor, where the $N_c$ color pairs
are analogues of $N$ Cooper pairs and the weak 4-fermion Fr\"ohlich Hamiltonian interaction is enhanced 
by a factor of $N_{Cooper}$ \cite{Cooper}.
The color factor enhancement also occurs in the NJL model, but at loop level.
Here we see that the color enhancement is occurring in the semi-classical (no loop) coloron 
theory by this coherent mechanism.  To our knowledge this semiclassical enhancement
of the underlying coupling strength in a coloron model, or its
approximation to QCD chiral dynamics, has not been previously noted.

Factorizing $\Phi^0$,
 \smbbo
\label{factor000}
\sqrt{J/2}\; \Phi^0(X,r)= \chi(X)\phi(r),
 \smebo
the kinetic term action becomes identical to the bosonic case,
\smbbo
\label{act1}
\backo\backo
S_K= \!\!\int_{X} \bl|\partial_X\chi(X)|^2-|\chi(X)|^2\int'_{\vvr}\!|\partial_{\vvr}\phi(\vvr)|^2
\br,
\smebo
with,
\smbbo
\int'_{\vvr}|\phi(\vvr)|^2=1.
\smebo
The interaction term
can then be written from eq.(\ref{42}) as,
\smbbo
\label{interaction00}
\backo
S_C
\rightarrow
g^2JM^4N_c\!\!\int_{X,r}\!\!\!\! \; \Phi^0{}^\dagger(X,r)\; D_F(2r)\;\Phi^0(X,r)
\nonumbo
=
g^2N_c\!\int_{X} |\chi(X)|^2 \int'_{\vvr}|\phi(\vvr)|^2 M\frac{e^{-2M|\vvr|}}{8\pi |\vvr|}.
\smebo
Hence, the removal of relative time 
is identical procedure as in the bosonic model,
and we obtain to the same action, $S=S_K+S_C$ in the compact notation of eqs.(\ref{compact1})
with the interaction now enhanced by $N_c$.

The extremalization of $\phi$ then leads to the 
SKG equation,
\smbbo
\label{acteqn2}
\backo
-\nabla_{\vvr}^2\phi(\vvr) -{g^2N_c M}\frac{e^{-2M|\vvr|}}{8\pi|\vvr|}\phi(\vvr)= m^2{\phi}(\vvr).
\smebo

\subsection{Classical Criticality of the Coloron Model }

The coloron model  furnishes a  direct UV
completion of the NJL model.  However, in the coloron model we do not
need to invoke large-$N_c$ quantum loops to have a critical theory.
Rather, it leads to an SKG potential of the Yukawa form
which has a {\em classical critical coupling}, $g_c$.
For $g<g_c$ the theory is subcritical and produces
resonant bound states that decay into chiral fermions.
For $g>g_c$ the theory produces a tachyonic bound state which implies
a chiral instability and $\Phi$ must develop a VEV. This
requires stabilization by, e.g., quartic interactions and a sombrero potential.
All of this is treated bosonically in our present formalism.

The criticality of the Yukawa potential in the nonrelativistic
Schr\"odinger equation is discussed in the literature in the context of ``screening.''
The nonrelativistic Schr\"odinger equation $r=|\vvr|$ is:
 \smbbo
-\nabla^2\psi - 2m\alpha\frac{e^{-\mu r}}{r}\psi=2mE
 \smebo
and criticality (eigenvalue $E=0$) occurs for $\mu=\mu_c$ where a numerical analysis yields,
\cite{Edwards},
 \smbbo
\mu_c= 1.19\;\alpha m.
 \smebo
For us the spherical SKG equation is now  eq.(\ref{acteqn2}).  
Comparing,  gives us a critical value of the coupling constant,
when $\mu_c\rightarrow 2M$, $m\rightarrow M/2$ and $\alpha\rightarrow g^2N_c/8\pi$, then:
 \smbbo
 \label{critc1}
\backo\!\!\!
2M=(1.19)\bl\!\frac{M}{2}\!\br\bl\!\frac{g^2N_c}{8\pi}\!\br, \;\;\makebox{hence:}\;\;
{g^2}/{4\pi}=
6.72/N_c.
 \smebo
We can compare the NJL critical value of eq.(\ref{NJL3}), 
\smbbo
\label{critc2}
g_{cNJL}^2/4\pi = 2\pi/N_c=6.28/N_c.
\smebo
Hence, the NJL quantum criticality is a comparable effect, with a remarkably
similar numerical value for the critical coupling.

Note that we can rewrite eq.(\ref{acteqn2}) with dimensionless coordinates, $\vec{u}=M \vvr$,
\smbbo
\label{crit4}
M^2\bl-\nabla_u^2\phi(u) -{g^2N_c}\frac{e^{-2u}}{8\pi u}\br\phi(u)= 0,
 \smebo
and then $M^2$ only appears as an overall scale factor.   Hence we see that critical coupling is
determined by eq.(\ref{crit4}), and the scale $M$ cancels out at criticality. 
The mass  scale $M$ is dictated
in the exponential $e^{-2Mr}$ of the Yukawa potential.  
However, in general, we can start with the dimensionless form of
the SKG equation, as in eq.(\ref{crit4}),
and infer the scale $M$ by matching to any desired potential. In this way, solutions may exist
where the scale in the potential is driven by the renormalization group,
as in the Coleman-Weinberg mechanism \cite{CW}\cite{CTHCW}. This will
be investigated elsewhere.

However, it is important to realize that the NJL model involves
the Yukawa coupling, $g_{NJL}$ while the present criticality involves the 
coloron
coupling constant.
The Yukawa coupling is emergent in the coloron model, and we need to compute it.

\subsection{Yukawa Interaction}

The second term in eq.(\ref{interaction00}) is the induced the Yukawa interaction $S_Y$,
and can be written as:
\smbbo
\label{Yukawa000}
S_Y=g^2(\sqrt{{2JN_c}})M^2\times
\nonumbo
\int_{Xr}\!\bar{\psi}^a_L(X-r)\psi_{aR}(X+r) D_F(2r) \chi(X) \phi(\vvr){+h.c.}
\smebo
This is the effective Yukawa interaction between
the bound state $\Phi^0$ and the free fermion scattering states.

We can't simply integrate out the relative time
here. 
However, we can first connect this to the point-like limit by suppressing
 the $q^2$ term in the denominator of $D(2r)$ with $z\rightarrow 0$ in:
 \smbbo
 \backo
 {D}_F(2r)\rightarrow \!\!\int\!\!\frac{1}{M^2}e^{2iq_\mu r^\mu}\frac{d^4q}{(2\pi)^4}
 \rightarrow \frac{1}{JM^2}\delta^4(r),
 \smebo
 where $\delta^4(2r)=J^{-1}\delta^4(r)$,
 hence with $J^{-1}=1/16$,
 \smbbo
 \label{74}
 \backo
 S_Y=g^2(\sqrt{{N_c}/{8}})
\int_{X}\!\bar{\psi}^a_L(X)\psi_{aR}(X) \chi(X) \phi(0){+h.c.}
 \smebo
 This gives the Yukawa coupling,
 \smbbo
 \label{gY}
 g_Y=g^2(\sqrt{{N_c}/{8}})\phi(0).
 \smebo
 The wave-function at the origin, $\phi(0)$ in the NJL limit
 is somewhat undefined.  However, if we consider a spherical cavity
 of radius $R$ where $MR=\pi/2$, with a confined,  dimensionless, $\phi(r)$,
 then $\phi(0)$ is obtained in eqs.(\ref{B57},\ref{B58})
\smbbo
\phi(0)= \frac{1}{\pi}.
\smebo
Plugging this into the expression for $g_Y$ in eq.(\ref{gY}) gives,
\smbbo
\label{77}
g_{Y}  =g^2\sqrt{{N_c}/{8\pi^2}}=\frac{g^2}{g_{cNJL}},
\smebo
where $g^2_{cNJL}=(8\pi^2/N_c)$ is the critical coupling of the
NJL model, as seen in eq.(\ref{NJL3}).

Hence, if the coloron coupling constant,  $g^2$,  is critical, as in eqs.(\ref{critc1},\ref{critc2}).
then we have seen in that $g^2\approx g^2_{cNJL}$,
and  the induced Yukawa coupling from eq.(\ref{77}) is then $g_Y\approx g_{cNJL}$.
The coloron model is then consistent with the NJL model
in the point-like limit where the NJL model coupling is the Yukawa coupling, as seen in eq.(\ref{NJL2}).   

However, the induced Yukawa coupling
in the bound state, $g_Y$, may be (significantly?) different than the coloron coupling $g$
in realistic extended $\vvr$ models.
The result we just obtained applies when we assume the strict  point-like limit of $D_F(2r)\sim \delta^3(r)$,
while in reality, as the potential becomes more extended,  the $\int V(\vvr)\phi(\vvr)$
may become smaller, even if
the coloron coupling $g$ may be supercritical.  We anticipate this could have implications for a composite
Higgs model, which will be investigated elsewhere (together with loop effects including the 
extended wave-function). 
There may also be additional new effects that occur at loop level in extended potentials,
such as the infall of zero modes, as suggested in \cite{cth0}.\footnote{In ref.(\cite{cth0})
we considered the original suggestion of Yukawa \cite{Yukawa} 
that assumes the interaction of eqs.(\ref{Yukawa00},\ref{Yukawa000})
with $D_F(2r)=1$ for $r<M^{-1}$. This yielded, from fermion loops, a $\sim -g^2N_c/r$ potential for $\phi(r)$.
$D_F(2r)=1$ requires a flat potential
for $r<M^{-1}$, but can produce a naturally low mass scalar. }

\subsection{Spontaneous Symmetry Breaking}

For subcritical
coupling there are resonance solutions with positive $m^2$ that have large distance tails of 
external incoming and outgoing radiation, representing a steady state
of resonant production and decay. 
We give such a solution (with a barrier potential) in Appendix \ref{barrier}.
The portion of the wave-function within the
potential can be viewed as the resonant bound state for normalization purposes,
while the large distance tail is non-normalizable
radiation. 

With super-critical coupling, $g>g_c$, the bilocal field $\Phi(X,r)$ has a negative squared mass
eigenvalue (tachyonic), with a well-defined localized wave-function.  In the region external to the
potential (forbidden zone)
the field is exponentially damped. At exact criticality with $g=g_c-\epsilon$ there is a $1/r$
(quasi-radiative) tail that switches to exponential damping for $g=g_c+\epsilon$.
The supercritical solutions are localized and normalizable over the
entire space $\vvr$, but with $m^2<0$ they lead to exponential
runaway in time of the field $\chi(X^0)$, and must be stabilized, typically
with a $|\Phi|^4$ interaction.

We then treat the supercritical case as resulting in spontaneous symmetry breaking. 
In the point-like  limit, $\Phi(X)\sim \Phi(X,0)$, 
the theory has the ``sombrero potential'', 
  \smbbo
 V(\Phi) = -|M^2\Phi^2|+\frac{\lambda}{2}|\Phi|^4.
  \smebo
The point-like field develops a VEV, $\langle \Phi \rangle =|M|/\sqrt{\lambda}$.
In this way the bound state theory will drive 
the usual chiral symmetry breaking
from the underlying dynamics of a potential induced by new physics.

If a quartic potential exists in a local field theory, eq.(\ref{start}),  we can bilocalize
it by introducing another world scalar, 
\smbbo
\backo
\frac{\lambda_0}{2}\int d^4x|\varphi|^4\rightarrow \frac{\lambda_0}{2}
\int d^4y \;Z'|\varphi(y)|^4\!\!\int d^4x|\varphi|^4
\nonumbo
\backo =Z'MT\frac{\hat\lambda}{2}\!\!
\int_{X}\!\!\int_r' \!\!|\chi(X)\phi(\vvr)|^4,
\smebo
and  $Z'MT=1$ to absorb relative time. 

The simplest sombrero potential can therefore be modeled
as,
 \smbbo
\backo\;
S=\!\!\int_{X\vvr}' \!\bl \!|\phi|^2\!\left|\frac{\partial\chi}{\partial X}\right|^2\!
\!\!-\!|\chi|^2( |\nabla_r\phi|^2\!
+\!g^2N_cM V(r)|\phi(r)|^2) 
\nonumbo
\qquad\qquad
-|\chi|^4\frac{\hat\lambda}{2}|\phi(\vvr)|^4\br.
 \smebo
In the case of a perturbatively small $\lambda$
we expect the eigensolution of $\phi$
to be essentially unaffected,
\smbbo
\int'_r \!\! \bl \!|\nabla_{\vvr}\phi|^2\!
+\!{g^2N_c\!M}V(\vvr)|\phi(\vvr)|^2\!\br \approx m^2.
\smebo
The effective quartic coupling is then further renormalized by
the internal wave-function,
\smbbo
\frac{\hat \lambda}{2}|\chi|^4\int'_r |\phi(\vvr)|^4=|\chi|^4\frac{\widetilde\lambda}{2}.
\smebo
In this case we
 see that $\chi$ develops a VEV in the usual way:
 \smbbo
\langle |\chi|^2 \rangle = |m^2|/\widetilde\lambda =v^2.
 \smebo
This is consequence of $\phi(\vvr)$ remaining localized in its potential

The external scattering state fermions, $\psi^a(X)$, will then acquire mass 
through the emergent
Yukawa interaction described in the previous section, $\sim g_Y\langle |\chi| \rangle.$
However, an issue we have yet to resolve is whether 
 the induced fermion masses back-react the VEV solution itself?
We segregated the free fermions from the bound state
wave-function, $\Phi$, by shifting, so we are presently arguing that
$\Phi$ forms a VEV  as described above, and the scattering state fermions
independently 
acquire mass as spectators, but this may require a more
detailed analysis.
 
 However, for general $\widetilde\lambda$ and possibly large
 (as in a nonlinear sigma model),
 the situation is potentially more complicated.
 The VEV is determined by joint integro-differential equations for constant $\chi$,
 \smbbo
 0= \!\!\int'_r \bl \!|\nabla_{\vvr}\phi|^2\!
+\!{g^2N_c\!M}V(r)|\phi(\vvr)|^2\!
+{\widetilde\lambda}|\chi|^2||\phi(\vvr)|^4 \br
\nonumbo
0=
-\nabla^2_r\phi\!+\!{g^2N_c\!M}V(r)\phi(r)
-{\widetilde\lambda}|\chi|^2|\phi(r)|^2\phi(r).
 \smebo
 If we can can solve the second local equation then the global one
 follows, but we see that
  $\phi(r)$ cannot
 become constant in a potential $V(r)$ which has $r$ dependence! 
 While perturbative 
 solutions maintain locality in $\phi(\vvr)$, it is unclear what solutions exist to
 the latter equation for non-perturbative $\lambda$.
 If $\phi(r)$ acquires a constant component then we may have something
 analogous to the Bose-Einstein condensate (BEC) phase of a superconductor.
 

\section{Summary and Conclusions}

In the present paper we have given a 
formulation of bilocal field theory, $\Phi(x,y)$, as a variation on Yukawa's original 
multilocal field theory of composite particles
 \cite{Yukawa}. In particular we focus upon  two  particle bound states consisting
 of bosons or chiral fermions. There are many foreseeable extension of the
 present work.
and scattering states

Here we construct bilocal field theories from an underlying 
local interacting field theory, by
introduction of ``world-scalars.'' We then go to barycentric coordinates, 
and the bilocal field is 
``factorized,''
\smbbo
X=\half(x+y),\qquad r=\half(x-y)
\nonumbo
\Phi(x,y)\rightarrow \Phi(X,r)=\chi(X)\phi(r).
\smebo
  Here
$\chi(X)$ describes center-of-mass motion like any pointlike
scalar field, while $\phi(r)$ is the internal wave-function of the bound state.

This procedure enables the removal of the relative time, $r^0$, 
in the bilocalized theory, essentially by canonical renormalization.   The bilocal kinetic term
contains a constraint that leads to a static internal field,  
$\phi(r)\rightarrow \phi(\vvr)$, in the center of mass frame.
Hence, we obtain a static Schr\"odinger-Klein-Gordon (SKG) equation for the internal wave-function.
The eigenvalue of this equation is the $m^2$ of the bound state.  

The SKG equation likewise contains a static potential that comes from 
the Feynman propagator of the exchanged particle
in the parent theory. Typically we expect, and have analyzed,
a Yukawa potential, $\sim -g^2 Me^{-2Mr}/8\pi r$, though the
formalism can in principle  accommodate any desired phenomenological potential.
Here $g$ is the exchanged particle
coupling constant, such as the coloron coupling (massive perturbative gluon) 
for fermions. This is not the scalar-fermion Yukawa coupling, $g_Y$, which is
subsequently emergent.  

We find that the Yukawa potential  is classically critical with coupling $g_c$. If 
the coupling is sub-critical, $g<g_c$,
then $m^2$ positive, and the bound state is therefore a resonance.
It will decay to its constituents if kinematically allowed.
$\phi(\vvr)$ is then a localized ``lump,'' with a radiative tail
representing the two body decay and production by external free particles
as in Appendix \ref{barrier}.

If the coupling is supercritical, $g>g_c$, then $m^2<0$, is tachyonic and $\Phi$ will acquire a VEV.
We require an interaction, such as $\sim \lambda|\Phi|^4$,
to stabilize the vacuum and we therefore have spontaneous
symmetry breaking.   $\phi(r)$ is expected to be localized in its potential and $\chi(X)$
acquires the VEV.  If $\phi(r)$ becomes delocalized; both $\chi(X)$
and $\phi(r)$ acquire VEVs, which is analogue of a Bose-Einstein condensate
e.g., in a slightly heated superconductor,
however we have not produced solutions to the SKG equation that demonstrates this behavior.

We consider a bound state of chiral fermions in the coloron model, where a coloron is a
massive gluon with coupling $g$, such as in ``topcolor'' models \cite{Topcolor}
and chiral constituent quark models \cite{bijnens}.
Fierz rearrangement of the non-local, color-current-current, interaction yields 
a leading large--$N_c$ interaction in
$\bar{\psi}(x)_L\psi(y)_R\sim \Phi(x,y)$. 
For the color singlet $\Phi$ the coupling $g^2$ is
enhanced by $N_c$, in analogy to a  BCS superconductor \cite{Cooper}.

As our main result, we find that the coloron model can be  classically critical.
The critical value of $g^2$,  extracted from \cite{Edwards}, is astonishingly close to the 
critical value of the Yukawa coupling in the NJL model.  In the NJL model the
critical behavior is ${\cal{O}}(\hbar)$ and comes 
from fermion loops. In the present model this is a semiclassical result, and the 
essential factor of $N_c$ comes
from the coherent BCS-like enhancement of the 4-fermion scattering amplitudes.
It is therefore unclear what happens when we include the fermion loops in addition to the  
classical behavior in the large $N_c$ limit. Is criticality further enhanced by the additional loop
contribution? Is there an additional $N_c$ enhancement of the underlying coupling due to the $\sqrt{N_c}$
factor in the emergent Yukawa interaction? These are interesting issues we will address elsewhere.

The induced Yukawa coupling of the bound state to fermions, $g_Y\bar{\psi}(x)_L\psi(y)_R \Phi(x,y)$,
is extended and emergent in the composite models.
We derive the  coupling and find 
 $g_Y\propto \int V(r) \phi(r) \sim g^2\phi(0)$ in the point-like limit.  
 For $\phi(0)$ in a tiny spherical cavity we obtain 
 $g_Y=g^2/g_{cNJL}$.  Hence the critical value of $g^2$ implies the 
critical value of $g_Y=g_{cNJL}$ in point-like limit NJL model, which
is consistent.

However,  the $\int V(r) \phi(r)$ could in principle be reduced in an extended potential, 
as the wave-function spreads out, even
if $g^2$ is critical.  
This is an intriguing possibility: the BEH-Yukawa coupling 
of the top quark is $g_{top}\sim 1$, which is perturbative and is insufficient
to drive the formation of a composite, negative $m^2$, Brout-Englert-Higgs (BEH) boson at low energies
in e.g., a top condensation model.  
However
the present result suggests that perhaps $\int V(r) \phi(r) $ is small
suppressing $g_{top}$, even though the underlying coloron coupling, $g^2N_c$, is
super-critical and leads to the composite BEH mechanism. In this picture
the BEH boson may be a large object, a ``balloon,'' of size $\sim m^{-1}_{top}$ (see  \cite{cth0}).

While we have an eye to a composite BEH boson for the standard model, as
in top condensation theories, \cite{Yama}\cite{BHL}\cite{CTH},
our present analysis is more general, and does not include details, e.g. gauge interactions
and gravity.
We plan to develop this approach further in future work.

Perhaps the key result in our formalism is the SKG
equation in scale invariant coordinates, $u = Mr$ of eq.(\ref{crit4}).  
This factorizes the mass scale $M$ into
an overall factor in the action, and the potential is then a scale-invariant function.  
This indicates that $M$ is ultimately
determined by the potential, and it can in principle arise from
the renormalization group (trace anomaly). For example, we can consider Coulombic binding of massless
fermions by introducing into eq.(\ref{crit4}) a potential, $\sim -g^2(u)/|\vec{u}|$, 
where $g^2(u)$ is determined
by the RG. We then invert the process of rescaling, $u\rightarrow Mr$,
and adjust $M$ to fit a physical $g^2(Mr)$, hence obtain the analogue of eq(\ref{crit000})
with a potential $\sim -M g^2(Mr)/|\vvr|$.   In this way the scale $M$ is arising by the boundary
conditions in the RG equation for a physical $g^2(Mr)$, in analogy
to the Coleman-Weinberg mechanism for dynamical symmetry breaking \cite{CW,CTHCW}.

While many elements of this formulation may be  implicit in the 
literature we think the emphasis here upon a bosonic field description, 
the  classical criticality of the coloron model, the  linkage to the
the NJL model by UV completion, and our treatment of relative time renormalization
comprise a novel perspective.   We will significantly expand upon this work elsewhere \cite{CTHprep}.

\appendix

\section{Notation $\label{SumNote}$}

Barycentric coordinates:
 \smbbo
\label{barycentric}
X=\half(x+y)\qquad r=\half(x-y)
\nonumbo
\partial_x=\half(\partial_X+\partial_r)\qquad
\partial_y=\half(\partial_X-\partial_r)
 \smebo

Two body scattering states:
 \smbbo
\Phi(x,y)=\exp(ip_1x+p_2y)=\exp(iP_\mu X^\mu+Q_\mu r^\mu)
\nonumbo
(\partial_x^2 +\partial_y^2)\Phi(x,y)=-(p_1^2+p_2^2)\Phi(x,y)=-2\mu^2\Phi(x,y)
\nonumbo
\bl \frac{\partial^2}{\partial X^\mu\partial X_\mu} 
+\frac{\partial^2}{\partial r^\mu\partial r_\mu}\br\Phi(X,r) =4\mu^2 
\nonumbo
(\partial_x^2 -\partial_y^2)\Phi(x,y)=
\frac{\partial^2}{\partial X^\mu\partial r_\mu}\Phi(X,r) =0
\nonumbo
dx^2+dy^2= 2dX^2+\half d\rho^2= 2dX^2+2dr^2
\nonumbo
\partial_x^2+\partial^2_y=\half\partial_X^2 +\half\partial r^2 
\qquad
\partial_x^2-\partial^2_y= \partial^\mu_X\partial_{\mu r} 
 \smebo

Integration Measures:
 \smbbo
 \int_{u...v;\vvx...\vvy}\!\!\!\! =\int\!\!d^4u..d^4v\;d^3x...d^3y
\nonumbo
 \int'_{u...v;\vvx...\vvy}\!\!\!\! =\int\!\!M^4d^4u..M^4d^4v\;M^3d^3x...M^3d^3y
\nonumbo
\int d^4xd^4y =J\!\! \int d^4X d^4r; 
\;\;\;
\makebox{Jacobian}\;\;J=(2)^4
\smebo

\section{Bilocal Field Theory $\label{4}$}
 
 Here we give a general discussion of our ``revisited'' bilocal field theory for a pair
 of particles, as inspired by Yukawa \cite{Yukawa}.  We begin with the ``bilocalization''
 and subsequently construct the generic actions for bilocal fields containing free particles.

\subsection{Free Fields}

Let us examine the bilocalization procedure in Section \ref{bloc} for 
free particle states.
Consider a pair of local scalar fields $\varphi_i(x)$ (complex):
 \smbbo 
 \label{A1}
\backo\backo
S=\int_x\bl |\partial\varphi_1|^2 + |\partial\varphi_1|^2 -\mu^2(|\varphi_1|^2+|\varphi_2|^2),
 \smebo
 with independent free particle equations of motion,
 \smbbo
 \partial^2\varphi_1+\mu^2\varphi_1=0\qquad \partial^2\varphi_2+\mu^2\varphi_2=0.
 \smebo
 We want to describe a pair of particles by a bilocal field of mass dimesion $1$:
 \smbbo
 \Phi(x,y) = M^{-1}\varphi_1(x)\varphi_2(y).
 \smebo
 We therefore have the two equations of motion for $\Phi$,
 \smbbo
 \label{eq1}
 (\partial_x^2+\partial_y^2)\Phi(x,y)+2\mu^2\Phi(x,y)=0
 \\ &&
 \label{eq2}
  (\partial_x^2-\partial_y^2)\Phi(x,y)=0.
 \smebo
 
 \subsection{Bilocalization of Scattering States}
 
 We can obtain the action for $\Phi$ as follows.
 We multiply the kinetic and mass terms of eq.(\ref{A1}) by
 world-scalars:
 \smbbo
 W_i=ZM^2\int\!\!d^4y\;|\varphi_i(y)|^2,
 \smebo
 hence,
 \smbbo 
\backo
S=\int_x\bl W_2|\partial\varphi_1|^2 +W_1|\partial\varphi_2|^2 -\mu^2(W_2|\varphi_1|^2+W_1|\varphi_2|^2)\br
\nonumbo
\backo
=ZM^2\!\!\int_{xy} \!\!\bl |\varphi_2(y)\partial_x\varphi_1(x)|^2
+|\varphi_1(y)\partial_x\varphi_2(x)|^2
\nonumbo \qquad \qquad -2\mu^2|\varphi_1(y)\varphi_2(x)|^2\br.
\smebo
At this stage we can still vary with respect to either $\varphi_1$
of $\varphi_2$, and the equations are modified.  In terms of the bilocal field we have,
\smbbo
\backo
=ZM^4\!\!\int_{xy}\!\! \bl |\partial_x\Phi(x,y)|^2
+|\partial_y\Phi(x,y)|^2 -2\mu^2|\Phi(x,y)|^2\br.
\nonumbo
 \smebo
We go to barycentric coordinates and note the derivatives:
 \smbbo
\label{barycentric}
X=(x+y)/2\qquad\qquad r=(x-y)/2
\nonumbo
\partial_x=(\partial_X+\partial_r)/2 
\qquad
\partial_y=(\partial_X-\partial_r)/2
\nonumbo
\Phi(x,y)\rightarrow \Phi(X,r).
 \smebo
The factor of $M$ is superfluous 
 at this point and in what follows we can set $M=1$ (we'll restore it below).
 Hence:
 \smbbo
 \backo \;\; S=
 \!JZ
 \!\!\int_{Xr} \!\!\bl\! \frac{1}{4}|(\partial_X+\partial_r)\Phi(X,r)|^2
+\frac{1}{4}|(\partial_X-\partial_r)\Phi(X,r)|^2 
\nonumbo 
-2\mu^2|\Phi(X,r)|^2\br,
\smebo
which yields the action,
\smbbo
\label{a1}
\backo\; S=\!
 \frac{JZ}{2}\!\!\int_{Xr} \!\!\bl |\partial_X\Phi(X,r)|^2
+|\partial_r\Phi(X,r)|^2 -4\mu^2|\Phi(X,r)|^2\br.
\nonumbo
 \smebo
 If we vary $\delta \Phi(x',y') = \delta^4(x-x')\delta^4(y-y')$
 we obtain one equation of motion:
 \smbbo
 \partial^2_X\Phi(X,r)+\partial^2_r\Phi(X,r)+4\mu^2\Phi(X,r)=0.
 \smebo
 We see that this is consistent with eq.(\ref{eq1}),
 \smbbo
 \partial_x^2+\partial_y^2 = \half(\partial^2_X+\partial^2_r).
 \smebo
However,  eq.(\ref{eq2}) is missing.  We see using,
 \smbbo
 (\partial_x^2-\partial^2_y)=\frac{\partial }{\partial X^\mu} \frac{\partial}{\partial r^\mu},
\smebo
 eq.(\ref{eq2}) takes the form,
 \smbbo
 \frac{\partial }{\partial X^\mu} \frac{\partial}{\partial r^\mu}\Phi(X,r)=0.
 \smebo
 As before, this can be viewed as a constraint, and we can treat it
 as a Lagrange multiplier as in eq.(\ref{LM}) to supply the second equation.
 
 We factorize $\Phi$ in barycentric coordinates as,
 \smbbo
\label{f1}
\sqrt{J/2}\; \Phi(X,r)= \chi(X)\phi(r).
 \smebo
 The action eq.(\ref{a1}) becomes,
 \smbbo
\label{a2}
\backo \backo S=Z
\!\!\int_{Xr} \!\!\bl |\phi(r)|^2|\partial_X\chi(X)|^2
\nonumbo \qquad
+|\chi(X)|^2(|\partial_r\phi(r)|^2 -4\mu^2|\phi(r)|^2)\br.
 \smebo
 We then define, 
 \smbbo
 1=Z\int d^4r|\phi(r)|^2.
 \smebo
 The Schr\"odinger-Klein-Gordon equation has eigenvalue $m^2$,
 \smbbo
 \backo
 \partial^2_{r} \phi(r) +4\mu^2\phi(r)=m^2\phi(r)
 \nonumbo
 \backo
 m^2\int_r|\phi(r)|^2= \int_r(|\partial_r\phi(r)|^2 +4\mu^2|\phi(r)|^2),
 \smebo
Then the $\chi$ equation becomes,
 \smbbo
 \partial^2_X \chi(X) +m^2\chi(X)=0.
 \smebo
 The constraint then takes the form,
  \smbbo
 \frac{\partial \chi(X)}{\partial X^\mu} \frac{\partial \phi(r)}{\partial r^\mu}=0.
 \smebo
In the barycentric (rest) frame we can choose $\chi$ to
have 4-momentum $P_\mu=(m,0,0,0)$.
The constraint becomes  $P_\mu \partial^\mu_r\phi(r) =0$,
Therefore $\phi$ becomes a {\em static function} of 
$r^\mu=(0,\vvr)$.
The 
Schr\"odinger-Klein-Gordon equation is then a static equation,
\smbbo
 -\nabla_{\vvr}^2 \phi(\vvr) +4\mu^2\phi(\vvr)=m^2\chi(\vvr).
 \smebo
We define the static internal wave-function $\phi(\vvr)$ to be 
dimensionless, $d=0$, and  we normalize the dimensionless static wave-function (restoring $M$),
\smbbo
\label{Bnorm}
(M^3)\int d^3\vvr\; |\phi(\vvr)|^2 = 1.
\smebo
 We then have,
 \smbbo
 \backo
 1=Z(M^4)\!\!\int_r|\phi(r)|^2=Z(M^4)\!\! \int dr^0 \int d^3\vvr |\phi(\vvr)|^2=ZT(M),
 \nonumbo
 \smebo
 where an integral over relative time
has appeared leading to an overall factor $ZT(M)=1$ in the action for $\chi$.
The action becomes two nested parts,
\smbbo
\label{a2}
\backo
S=\!\!\int_{X} \!\!\bl |\partial_X\chi(X)|^2 - m^2|\chi(X)|^2\br
\nonumbo
\backo
m^2\!\!\int_{\vvr}\!\!  d^3\vvr\; |\phi(\vvr)|^2= 
\int_{\vvr} (|\nabla_{\vvr}\phi(r)|^2 +4\mu^2|\phi(r)|^2),
 \smebo
 with eq.(\ref{Bnorm}).

\subsection{Kinematics of Scattering States  $\label{Free}$}

We can more directly construct the bilocal fields, without appealing to world scalars, by considering
simple free particle kinematics.
For free particles of 4-momenta $p_i$ we see that $\Phi$ describes a scattering state,
 \smbbo
\Phi(x,y)
\sim \exp(ip_{1\mu} x^\mu+ip_{2\mu} y^\mu)
=\exp(iP_\mu X^\mu + i Q_\mu r^\mu),
\nonumbo
 \smebo
where,
 \smbbo
P_\mu = p_{1\mu}+p_{2\mu} \qquad Q_\mu= p_{1\mu}-p_{2\mu}.
 \smebo
For massive particles, $p_1^2=p_2^2=\mu^2$, $\Phi$  satisfies,
 \smbbo
 \backo\backo\;\;\;\;
\label{dyn0}
\bl \frac{\partial^2}{\partial X^\mu\partial X_\mu} 
\!\!+\!\!\frac{\partial^2}{\partial r^\mu\partial r_\mu}\!\!+\!4\mu^2\br\Phi(X,r) =0 .
\smebo
The constraint is the second equation,
\smbbo
\label{cons}
\backo\backo\;\;\;\;
(\partial_x^2 -\partial_y^2)\Phi(x,y)=
\frac{\partial^2}{\partial X^\mu\partial r_\mu}\Phi(X,r) =0,
 \smebo
and forces and forces $\Phi(X,\vvr)$ to have $r^0=0$ in the center of mass frame,
hence $P_\mu Q^\mu=0$ and $Q=(0,\vvq)$.

\vspace{3mm}
\noindent {\bf Constant Positive $m_0^2=4\mu^2$;  (Unparticle)}:
\vspace{3mm}

In the center of mass frame, $p_1^0=p_2^0$, $Q^0=0$, 
 \smbbo
Q^2=(p_1-p_2)^2=-(\vec{p}_1-\vec{p}_2)^2=-\vec{q}^{\;2}.
 \smebo
Hence, from eqs.(\ref{dyn0},\ref{cons}),
 \smbbo
P^2=4\mu^2+\vec{q}^{\;2}
\nonumbo
 \frac{\partial^2}{\partial X^\mu\partial X_\mu}\Phi(X,\vvr)= (m_0^2+\vec{q}^{\;2})\Phi(X,\vvr).
 \smebo
Therefore, if we were to try to interpret $\Phi$
as a ``bound state'' we see that is has continuum of
invariant ``masses'' $m^2= m_0^2+\vec{q}^{\;2}$.  This is evidently
an ``unparticle'' \cite{Georgi}. This implies it is not localized
and is a scattering state of asymptotic free particles each of mass $\mu=m_0/2$.

With the constraint of eq.(\ref{cons}) we can choose
$r^\mu=(0, \vvr)$ and $X^\mu = (X^0,0)$.
We factorize $\Phi=\chi(X)\phi(r)$ and the
factor field $\phi(r)$ then
satisfies the static SKG equation with eigenvalue $M^2$,
 \smbbo
-\nabla_{\vvr}^2\phi(\vvr) +m_0^2\phi(\vvr)=m^2\phi(\vvr),
 \smebo
and the solutions of the factorized field $\phi(r)$ are static, box normalized, plane waves,
with eigenvalues $ m^2\equiv m^2_{\vvq} = m_0^2+\vvq^{\;2}$,
 \smbbo
\phi(\vvr) = \frac{1}{\sqrt{V}}\exp (i\vvq\cdot \vvr  )  \qquad m^2_{\vvq} = \vvq^{\;2}+m_0^2.
 \smebo
$\chi(X)$ then satisfies the KG equation with $X^0=t$ and $\vvX=0$,
 \smbbo
\partial_t^2 \chi(t)+m^2_{\vvq}\chi(t)=0\qquad t=X^0,
 \smebo
and the $\chi(X) $ solution becomes, 
 \smbbo
\label{spectrum}
\chi(X)\propto \exp(im_{\vvq}t).
 \smebo
This is just the spectrum of the two body states of $\mu$-massive particles in
the barycentric frame.

\vspace{3mm}
\noindent {\bf Constant Negative $m_0^2$;  (Tachyonic Unparticle)}:
\vspace{3mm}

If we suppose $m_0^2<0$ then  the solutions for $\Phi(t, r)$
for small $\vvq$ are runaway exponentials, $\Phi\sim \exp(|m_0|t)$. 
The  eigenvalues are,
 \smbbo
 \qquad m^2_{\vvq} = \vvq{}^{\;2}-|m_0|^2.
 \smebo
This  analogous to a ``Dirac sea'' of negative mass
eigenvalues extending from $\vvq^{\;2}=0$
to $\vvq^{\;2}=|m_0|^2$, where the lowest mass state has $\vvq=0$
and eigenvalue $-|m_0|^2$.

This would therefore be
a tachyonic instability of unparticles, and  we then require higher field theoretic
interactions, such as $\lambda|\Phi|^4$, to stabilize the vacuum. 
This represents a spontaneous breaking of the chiral symmetry
and the constituent fermions will dynamically acquire masses
and the instability is halted.

However, a constant negative $m_0^2$ does not give a Higgs-like spontaneous symmetry
breaking mechanism.  If the potential is,
 \smbbo
V=-m_9^2|\Phi|^2+\frac{\lambda}{2}|\Phi|^4,
 \smebo
 (where we have constant $-m_0^2$ rather than a localized potential) 
then indeed the field develops a constant VEV:
 \smbbo
\langle \Phi \rangle = V + \Phi^0\qquad m_0^2= v^2\lambda.
 \smebo
Then $\Phi^0$ indeed becomes a massive BEH mode,
but with the continuous unparticle spectrum $m^2= |m_0|^2 + \vvq^{\;2}$.
To have an acceptable BEH mechanism, with a well--defined BEH boson, we therefore require a localized potential in $\vvr$
where $\langle \Phi \rangle =\langle \chi \rangle \phi(\vvr)$
and $\phi(\vvr)$ is a
localized eigenmode.

\vspace{3mm}
\noindent {\bf Removal of Relative Time and Generic Potential}:
\vspace{3mm}

A  point-like interaction in the underlying theory
interaction may produce an effective action
with a potential term that is a function of $r^\mu$, 
\smbbo
\label{23}
\backo\backo
S=
\frac{JM^4}{2}\!
\!\int_{Xr}\!\!\!(
Z|\partial_X\Phi|^2+ Z|\partial_r\Phi|^2 \!
-U(2r^\mu)|\Phi(X,r)|^2),
\smebo
 and we assume the constraint,
  \smbbo
\backo\backo
S =  \!\! \int_{Xr}  \eta\bl
\frac{\partial\Phi^\dagger }{\partial X^\mu} \frac{\partial\Phi}{\partial r_\mu}+h.c.\br^2.
 \smebo
We factorize $\Phi$,
 \smbbo
\label{factor000}
\sqrt{J/2}\; \Phi(X,r)= \chi(X)\phi(r).
 \smebo
In the center of mass frame we 
impose the constraint,
$\Phi(X,r)\rightarrow \chi(X^0)\phi(\vvr)$.
We can then integrate over relative time,
 \smbbo
\backo\backo
S =  
M^3\!\! \int_{X\vvr}  \bl
ZTM|\partial_X \chi|^2|\phi(\vvr)|^2
-ZTM|\chi(X)|^2|\nabla_{\vvr} \phi|^2 
\nonumbo
- V(\vvr) |\chi(X)|^2|\phi(\vvr)|^2\br,
 \smebo
 where,
 \smbbo
 V(\vvr)=M\int dr^0 U(2r^\mu),
 \smebo
 and $V(\vvr)$ has dimensions of $M^2$.
 We now define the normalizations, 
  \smbbo
  \label{Bnorm2}
 1=M^3 \!\! \int \!\! d^3r |\phi(\vvr)|^2\qquad\qquad  ZTM=1.
 \smebo
 The purpose of these normalizations is to have a canonically normalized $\chi$
 field in the center of mass frame, even in the limit $g=0$. It also defines
 the conserved $\chi$ current to have unit charge (one bound state pair)
 to match the underlying theory's conserved current.
 
 Hence,
\smbbo
\label{a2}
\backo
S=\!\!\int_{X} \!\!\bl |\partial_X\chi(X)|
-m^2|\chi(X)|^2\br
\nonumbo\backo
m^2\!\!\int_{\vvr}\!\!  d^3\vvr\; |\phi(\vvr)|^2= \int_{\vvr}(|\nabla_{\vvr}\phi(r)|^2 +V(r)|\phi(r)|^2),
 \smebo
 with eq.(\ref{Bnorm2}).

An attractive (repulsive) potential is $V<0$ ($V>0$).
The main point is that the field $\chi(X)$ must have canonical normalization
of its kinetic term, which dictates the introduction of $Z$.  The kinetic term
of $\chi$ is seen to be extensive in $T$, and this factor is absorbed by 
normalizing $Z$, as $ZMT=1$.
The potential
is determined by the integral over relative time of the interaction
and is not extensive in $T$.

 
From the factorization $\Phi=\chi(X)\phi(\vvr)$ with normalized $\phi(r)$, as in eq.(\ref{norm01}),
the eigenvalue $m^2$ is computed in the rest frame,
 \smbbo
 \backo\backo
\label{m2}
m^2= m^2 \int_{\vvr}|\phi(\vvr)|^2 =
\!\!
\int_{\vvr} \bl
-\phi^\dagger {\nabla^2 }\phi
+ V(\vvr)|\phi(\vvr)|^2 \br.
 \smebo
 From this we obtain
a normalized effective action for $\chi$ in any frame,
 \smbbo
\label{sep2}
S_\chi =
\int d^4X \bl
\frac{\partial \chi^\dagger }{\partial X^\mu}\frac{\partial \chi }{\partial X_\mu}
-m^2|\chi|^2\br.
 \smebo
The  mass of the bound state is determined by the
eigenvalue of the static Schr\"odinger-Klein-Gordon (SKG) equation:
 \smbbo
\label{SKG}
-{\nabla^2 }\phi+ V(\vvr)\phi(\vvr) =m^2\phi(\vvr).
 \smebo

\begin{figure}
	\centering
	\includegraphics[width=0.9\textwidth]{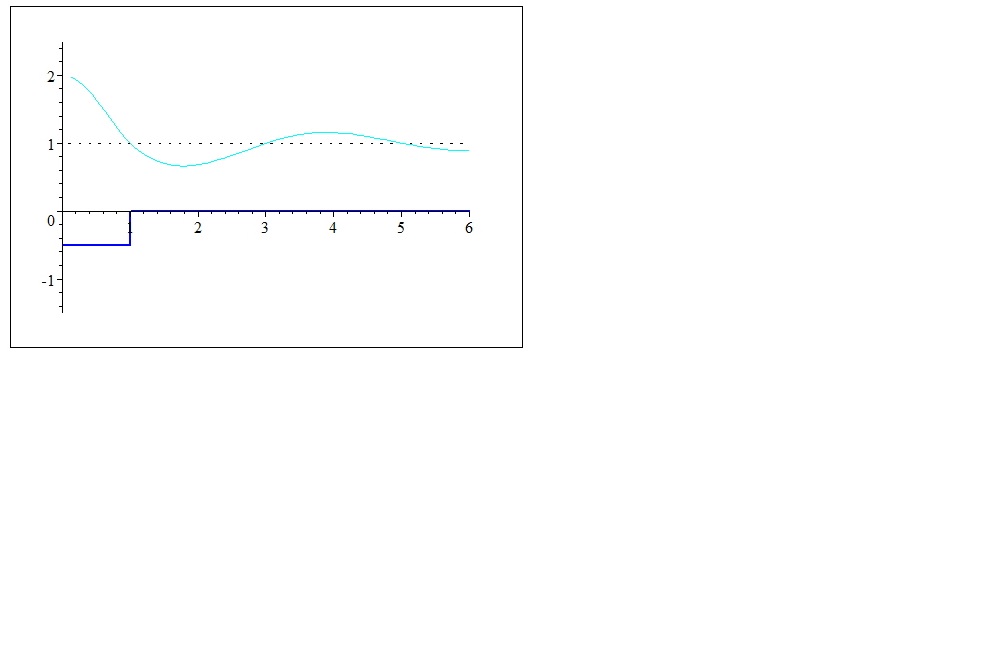}
	\vspace{-2.0in}
	\caption{ 
	Sub-critical $\phi(r)=u(r)/r$ (shallow potential, $-g^2R_0^2=-0.5$, arbitrary units,
	solution is schematic and unnormalized). The groundstate is a resonance
	with $m^2>0$ and is unstable, and is mostly defined by the region $r<0$
	corresponding to the maximum of the Lorentz line-shape.
	The external $r>1$ solution is a steady-state combination of incoming and outgoing radiation
	in the form of a Yukawa wave-function for a scattering pair (the decay width is estimated in
	analogy to the barrier potential below).  This would be the starting
	point for a Nambu-Jona-Lasinio model in which fermionic quantum loops pull $m^2$ negative.
	}
	\label{subcrit}
\end{figure}

\begin{figure}
	\centering
	\includegraphics[width=0.9\textwidth]{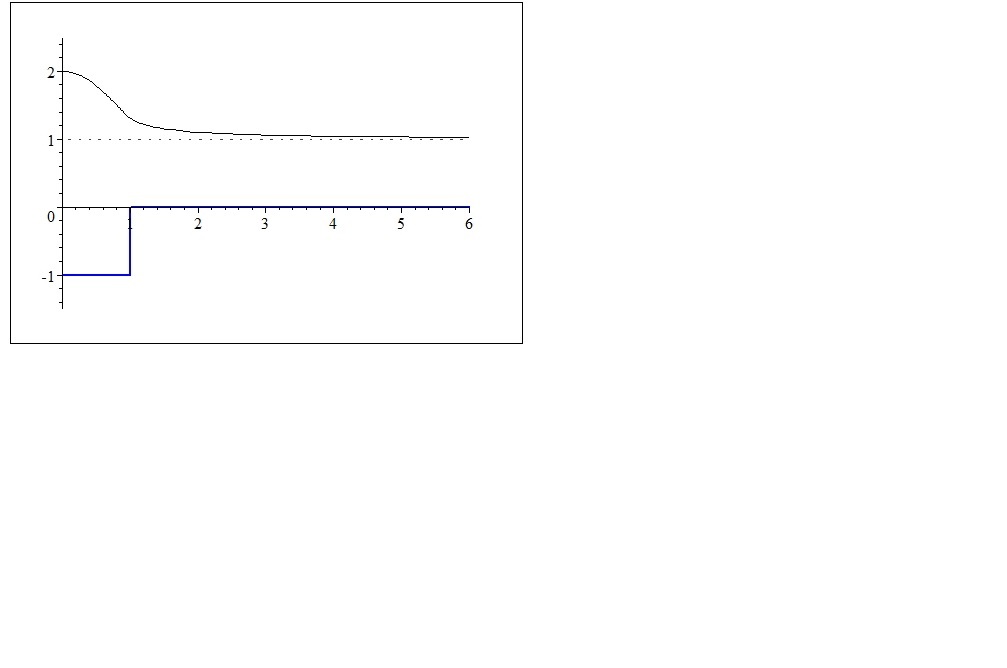}
	\vspace{-2.0in}
	\caption{ 
	Critical $\phi(r)=u(r)/r$ (medium depth potential, $-g^2R_0^2=-1.0$, arbitrary units,
	solution is schematic and unnormalized).  External solution,
	$r >1$, is a zero-mode $\phi(r)\propto 1/r$ and implies eigenvalue $m^2=0$. 
	}
	\label{crit}
\end{figure}

\begin{figure}
	\centering
	\includegraphics[width=0.9\textwidth]{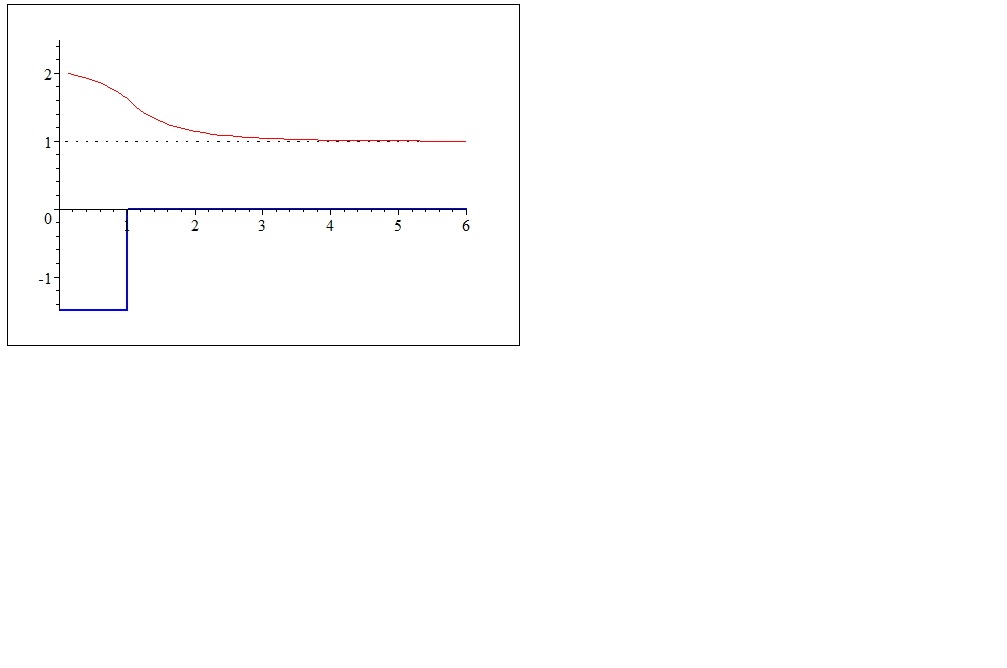}
	\vspace{-2.0in}
	\caption{ 
	Super-critical $\phi(r)=u(r)/r$  (deep potential, $-g^2R_0^2=-1.5$, arbitrary units,
	solution is schematic and unnormalized).  External solution,
	$r >1$ is exponentially damped into the forbidden region.  Solution has eigenvalue $m^2<0$ 
	and implies $\Phi(X,r)$ is tachyonic and will acquire a VEV and break
	chiral symmetry. This is an example of the classical (non-loop) potential initiating the
	chiral instability.
	}
	\label{supcrit}
\end{figure}

\vspace{3mm}
\noindent {\bf Spherical Potential}:
\vspace{3mm}

Consider the spherical potential and 
ground state,  where,
 \smbbo
\nabla^2\phi(r)=\frac{\partial^2}{\partial r^2}\phi(r)  + \frac{2}{r}\frac{\partial}{\partial r}\phi(r),
 \smebo 
where $r$ is the radius. 
Hence we use the usual reduction
$\phi(r)=u(r)/r$, where,
 \smbbo
\nabla^2\phi(r)= \frac{1}{r}\frac{\partial}{\partial r}u(r),
 \smebo
and the static SKG equation for the spherical state then becomes:
 \smbbo
\label{SKG20}
-\frac{\partial^2 }{\partial r^2}u(r)+  V_0(r)u(r) =m^2u(r),
 \smebo
with eigenvalue $m^2$. The $\chi(X)$ factor field then
satisfies the usual Klein-Gordon equation in $X^\mu$,
 \smbbo
\frac{\partial^2}{\partial X^\mu\partial X_\mu}  \chi(X) +m^2\chi(X)=0,
 \smebo
with 
plane wave solution, $\chi = \exp(iP_\mu X^\mu)$ and $P^2=M^2$.
The barycentric frame has $P^\mu = (m,0,0,0)$
$\chi = \exp(im X^0)$, and $m$ is determined by eq.(\ref{SKG20}).

The spherical potential well is defined by, 
\bea
\label{potwell}
\!V_0(r) &=& -g^2 M^2 \qquad r < R=\pi/2M \;\;\; \makebox{region I}
\nonumber \\
W_0(r) & =& 0 \qquad \qquad\;\;\; r > R=\pi/2M \;\;\;  \makebox{region II }.
 \smebo
 where we choose $R=\pi/2M$ and
have the two parameters $M$ and
$g^2$.  For fixed $M$ the depth of the potential is
controlled by $g$ and there is a critical value $g_c$.
For $g<g_c$ then $m^2>0$ and the solution is  a resonance as in Fig.(\ref{subcrit})
with a radiative tail representing incoming and outgoing radiation. For $g>g_c$
we have $m^2<0$ (tachyonic) 
and implies a vacuum instability and spontaneous symmetry breaking,  Fig.(\ref{supcrit}).
For $g=g_c$ the solution has zero eigenvalues and a scale invariant tail
$\propto 1/r$ Fig.(\ref{crit}).

The eigenfunctions are textbook, and we only summarize
their general structure. We take $u(r)=N\sin(kr)$, and
a negative eigenvalue (tachyonic) $m^2$ 
solution would take the form,
\bea
\label{k0}
\backo\!\!\!
u(r) & = & N\sin(kr) \qquad  m^2=k^2-g^2M^{2}   \qquad \makebox{region I.}
\nonumber \\
\backo\!\!\!
u(r) & = & \alpha e^{-\kappa (r-R)} \qquad\;  m^2=-\kappa^2   \qquad\qquad \makebox{region II.}
 \smebo
Matching  the field and its derivative at the region boundary $R_0$,
we have ,
 \smbbo
\cot(kR)=-\frac{\kappa}{k}\qquad m^2=-\kappa^2=k^2-g^2M^2.
 \smebo
The critical coupling $g_c$ is then determined by $m=\kappa=0$
hence $ kR=\pi/2$, hence $k=M$ and $g_c=1$,
\smbbo
\label{B56}
u(r) = \sin(Mr)\qquad \phi=N\frac{\sin(Mr)}{Mr},
\smebo
and the normalization is,
\smbbo
\label{B57}
1=N^2M^3\int d^3r \frac{\sin(Mr)^2}{M^2r^2}
={\pi^2} N^2,
\smebo
hence the field value at the origin is,
\smbbo
\label{B58}
\phi(0)= \frac{1}{\pi}.
\smebo



\section{A Simple Composite Resonance With Barrier Potential $\label{barrier}$ }

We illustrate a positive $m^2$ resonance, with its radiative tail of
incoming and outgoing radiation. We compute the decay width and find it matches
to a field theory calculation.
Consider a  barrier potential:
\bea
\label{barrier1}
\makebox{Region I:} &\qquad&  V_r(r<R)=0  \nonumber \\
\makebox{Region II:} &\qquad&   V_r(R<{r}<R+a)=W^{2}\nonumber \\
\makebox{Region III:} &\qquad&   V_r(r>R+a)= 0,
\eea
where 
the barrier, $W^2$, 
has dimensions of (mass)$^2$.
The general solution for eq.(\ref{SKG20}) is:
\bea
\makebox{Region I:} &\;\;&  \phi (r) = {{N}}\frac{\sin (k{r})}{r};\qquad   \nonumber \\
\makebox{Region II:} &\;\;&  \phi (r)=a\frac{e^{-\kappa(r-R)}}{r}
+b\frac{e^{\kappa(r-R)}}{r};\qquad 
 \nonumber \\
\makebox{Region III:} &\;\;\;& \phi (r)= \alpha \frac{e^{ik{r}}}{r}+\beta 
\frac{e^{-i{r}}}{r},
\eea
where $ \kappa =|\sqrt{W^{2}-k^{2}}|$. 

The solution represents a steady state, a balance of an incoming $\sim \alpha$
and outgoing $\sim \beta$ radiation in region III. Here the radiation appears
as two body, bilocalized states, as a linear combination of 
left-movers  $\sim r+t$ and right-movers  $\sim r-t$. 
The eigenvalue, $m^2$, is determined
by the incident radiation energy $k=m$.
To determine $N$ for a given  $(\alpha, \beta, k)$ we can use transfer matrices.
\vspace{0.1in}

\noindent
We match field value and derivative at the boundary of I-II:
\smbbo
\backo \backo \backo \backo\backo
M_{1}\left( 
\begin{array}{c}
a \\ 
b%
\end{array}%
\right) =M_{0}\left(\!\! 
\begin{array}{c}
N\sin kR \\ 
N\cos kR%
\end{array}%
\right) 
\nonumbo
\backo\backo \backo\backo\backo
M_{1}=\ \left( 
\begin{array}{cc}
\!\!\! 1 & \! 1 \\ 
\!\!\! -\kappa  & \! \kappa 
\end{array}%
\right) \qquad M_{0}=\ \left( 
\begin{array}{cc}
1 & 0 \\ 
0 & k%
\end{array}%
\right),
\smebo
\noindent
and match field value and derivative at the boundary of II-III:
 \smbbo
\backo\backo
M_{3}\left( 
\begin{array}{c}
\alpha  \\ 
\beta 
\end{array}%
\right) =M_{2}\left( 
\begin{array}{c}
\!\!e^{-\kappa a}a \\ 
\!\!e^{\kappa a}b%
\end{array}%
\right) =M_{2}P\left( 
\begin{array}{c}
a \\ 
b%
\end{array}%
\right)
\nonumbo
\backo\backo
M_{3}=\left( \ 
\begin{array}{cc}
\!\!\! 1 & \!\! 1 \\ 
\!\!\! ik & \!\! -ik%
\end{array}%
\right) \;\; M_{2}=\left( 
\begin{array}{cc}
\!\! 1 & \!  0 \\ 
\!\!  0 & \! \kappa 
\end{array}%
\right) \;\; P=\left( 
\begin{array}{cc}
\!\!\! e^{-\kappa a} & \!\!\! 0 \\ 
\!\!\! 0 & \!\!\! e^{\kappa a}%
\end{array}%
\right) .
\smebo
Therefore,
\smbbo
\left( 
\begin{array}{c}
\alpha  \\ 
\beta 
\end{array}%
\right) =M_{3}^{-1}M_{2}PM_{1}^{-1}M_{0}\left( \allowbreak 
\begin{array}{c}
N\sin kR \\ 
N\cos kR%
\end{array}%
\right),
\smebo
and we have,
\smbbo
M_{3}^{-1}M_{2}PM_{1}^{-1}M_{0}=\frac{1}{4}\left( 
\begin{array}{cc}
e^{-\kappa a}-\frac{i}{\epsilon }e^{\kappa a} &\;\;  -\epsilon e^{-\kappa
a}-ie^{\kappa a} \\ 
e^{-\kappa a}+\frac{i}{\epsilon }e^{\kappa a} &\;\;  -\epsilon e^{-\kappa
a}+ie^{\kappa a}
\end{array}%
\right)
\nonumbo
\epsilon =\frac{k}{\kappa }.
\smebo
One can readily check that $|\alpha|=|\beta|$.

Define  $F=\left| \alpha \right| ^{2}+\left| \beta \right|
^{2}$ and we have the result:
\smbbo
\backo
\!\!\!\!\!\!\!\!\!
F\!=\! \frac{N^{2}}{8\epsilon ^{2}}\!\bl\!\! e^{\!2\kappa a}(
\sin kR \!\!+\!\epsilon\! \cos kR) ^{2}\!
+ \!\!\frac{\epsilon^{2}}{e^{2\kappa
a}}\!
(\epsilon\! \cos kR \!-\!\sin kR)^{2}\!\!\br.
\smebo
The resonance is determined by maximizing $N^2$ for a given $F$,
or, for large $e^{\!2\kappa a}$ we can alternatively minimize $F$ for fixed $N$
and we see this corresponds to:
 \smbbo
 \tan kR+\epsilon =0\qquad \tan(kR)=-k/\kappa .
\smebo
Therefore, on resonance, 
\smbbo
\alpha=\beta=\half e^{-\kappa a} N \sin(kR)
\nonumbo
\left| \alpha \right| ^{2}+\left| \beta \right| ^{2}=\allowbreak \frac{k^{2}%
}{2\kappa ^{2}}N^{2}\left( \cos ^{2}kR\right) e^{-2\kappa a}.
\smebo
One can see the line shape by plotting as a function of $k$:
\smbbo
N^{2} \propto 
\frac{1}{
\left( \sin kR+\left( \cos kR\right) \epsilon \right) ^{2} } \qquad \epsilon=-k/\kappa
\smebo
The eigenvalue for the (mass)$^2$ is therefore $m^2=k^2$
and thus determined by $k$. 

Region III is radiation of the fermion pair, since for $\Phi=\chi\phi$, we  have:
\bea
\label{chisoln}
\!\!\!\!
\makebox{Region III:} &\;\;&
\Phi= \alpha\chi_0 \frac{e^{ik(t+r)}}{r}+\beta \chi_0
\frac{e^{ik(t-r)}}{r},
\eea
a sum of incoming (left-moving) and outgoing (right-moving) spherical waves.

\begin{figure}
	\centering
	\includegraphics[width=0.95\textwidth]{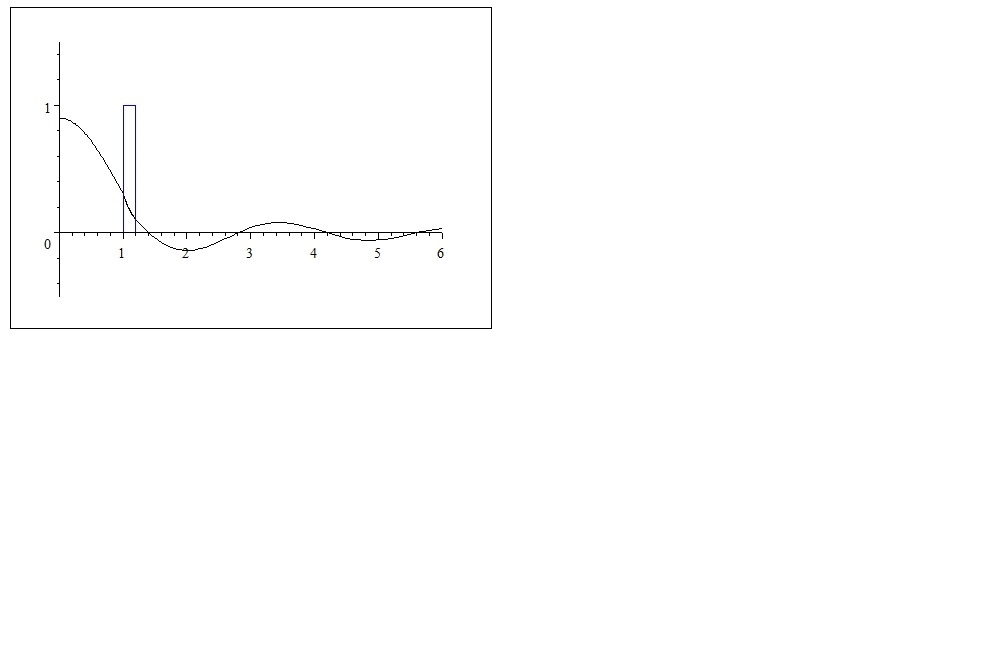}
	\vspace{-2.5 in}
	\caption{
	Barrier potential solution has $M^2>0$ and is an unstable resonance
	with external $r>1$ radiation. 
	This is also a candidate starting point for the Nambu-Jona-Lasinio model.
	}
	\label{fig:looper}
\end{figure}

The normalization of the resonance wave-function is defined by the cavity Region I
and yields approximately, with $kR=mR=\pi$:
\bea
\label{norm2}
&&
1=\int_r' | \phi^2|
=4\pi m^3 \int_0^R \!\!\!{{N}}^2\sin^2(m{r}) {dr}
={2\pi^2m^2{{N}}^2}
\nonumber \\
&&
{{N}}^2=\frac{1}{2\pi^2m^2}.
\eea
(note, the Region II contribution to the mass, in the large $\kappa$ 
and thin wall $a<<R$ limit, is
negligible
$\sim m^{2}a+O(am/\kappa ).$)

To compute the decay width of the resonant state, consisting of one pair
of fermions localized in the
Regions I+II,  we can switch off the incoming
radiation, $\alpha\rightarrow 0$, and the state will decay with
decay amplitude is $\beta.$  
The decay width is obtained semi-classically by the rate of energy
loss (power) into the outgoing spherical wave, divided by the mass.

Here's a semi-classical calculation of the decay width.
In the rest-frame with no explicit dependence upon $\vec{X}$
we see that $\Phi$ can be viewed as a localized field in $\vec{r}$ and $X^0=t$, hence
$\Phi(r,t)=\phi(r)e^{imt}/\sqrt{2mV}$, where $V$ is a volume.
The outgoing power is given by the stress tensor, $T_{0r}$,
from the right-mover solution and,
\bea
|\phi(r)|^2\sim |\beta|^2/r^2\sim {{N}}^2e^{-2\kappa a}/4r^2.
\eea
Hence the power, using eq.(\ref{norm2}):
\bea
P &=&  m^3
\!\!\int\!\! d^3 X {4\pi r^{2}} \left( \partial _{0}\Phi ^{\ast }
\partial_{r}\Phi  \right)\big|_{r\rightarrow \infty}
\nonumber \\
&\approx&  \pi m^4 V \frac{{N}^2}{(2V)} e^{-2\kappa a}=
\frac{1}{4\pi} m^2e^{-2\kappa a},
\eea
and the decay width is obtained as the ratio power to mass,
\bea
\Gamma &=&\frac{P}{m}= \frac{m}{4\pi}e^{-2\kappa a}.
\eea
We can compare the decay width from a complex point-like
field consisting on a single color $N_c=1$, of mass $m$ with Yukawa coupling $g$
to the fermions:
\bea
\Gamma =\frac{%
g^{2}}{16\pi }m,\qquad \makebox{hence,}\qquad
g=2e^{-\kappa a}.
\eea
We therefore have a heavy bound state with mass $M=k\approx \pi /R$
and an induced Yukawa coupling $g\propto e^{-\kappa a}$ which  is
perturbative
in the large $\kappa a\sim Wa$ limit. 

We've done this for a single color. In this simple model if
we extend to $N_c$ colors, then the coupling will be enhanced
by a factor of $N_c$ in analogy to the BCS theory,
\cite{Cooper}, as described in section $\ref{BCS}$. The
decay width is then,
\bea
\Gamma \rightarrow \frac{%
g^{2}N_c}{16\pi }m.
\eea
The usual field theory decay width with $N_c$
colors gives the same result in $N_c$.


\section*{Acknowledgments}
I thank W. Bardeen and Bogdan Dobrescu for critical comments,
and
the  Fermi Research Alliance, LLC under Contract No.~DE-AC02-07CH11359 
with the U.S.~Department of Energy, 
Office of Science, Office of High Energy Physics.

\end{document}